\documentclass[12pt,preprint]{aastex}
\begin{document}

\title{The Mid-Infrared Colors of the ISM and Extended Sources at the Galactic Center}

\author{
R.~G.~Arendt\altaffilmark{1,2}, 
S.~R.~Stolovy\altaffilmark{3}, 
S.~V.~Ram\'{\i}rez\altaffilmark{4}, 
K.~Sellgren\altaffilmark{5}, 
A.~S.~Cotera\altaffilmark{6}, 
C.~J.~Law\altaffilmark{7,8}, 
F.~Yusef-Zadeh\altaffilmark{7},
H.~A.~Smith\altaffilmark{9},
D.~Y.~Gezari\altaffilmark{10}
}

%\email {richard.g.arendt@nasa.gov}

\altaffiltext{1}
{CRESST/UMBC/GSFC, Code 665, NASA/Goddard Space Flight Center,
8800 Greenbelt Road, Greenbelt, MD 20771; richard.g.arendt@nasa.gov}
\altaffiltext{2}{Science Systems and Applications, Inc.}
\altaffiltext{3}
{{\it Spitzer} Science Center, California Institute of Technology, 
Mail Code 314-6, 1200 East California Boulevard, Pasadena, CA 91125}
\altaffiltext{4}
{IPAC, California Institute of Technology,
Mail Code 100-22, 1200 East California Boulevard, Pasadena, CA 91125}
\altaffiltext{5}
{Department of Astronomy, Ohio State University, 140 West 18th Av.,
Columbus, OH, 43210, USA}
\altaffiltext{6}
{SETI Institute, 515 North Whisman Road, Mountain View, CA 94043}
\altaffiltext{7}
{Department of Physics and Astronomy, Northwestern University,
Evanston, IL 60208}
\altaffiltext{8}
{Astronomical Institute ``Anton Pannekoek'', University of Amsterdam, 1098 SJ Amsterdam, Netherlands}
\altaffiltext{9}
{Harvard-Smithsonian Center for Astrophysics, 
60 Garden Street, Cambridge, MA 02138}
\altaffiltext{10}
{NASA/Goddard Space Flight Center, Code 667, 8800 Greenbelt Road, Greenbelt, MD 20771}

\begin{abstract}
A mid-infrared (3.6 -- 8 $\micron$) survey of the Galactic Center has been carried out 
with the IRAC instrument on the {\it Spitzer Space Telescope}. This survey
covers the central $2\arcdeg \times1\fdg4$ ($\sim 280\times 200$ pc) of the Galaxy. 
At 3.6 and 4.5 $\micron$ the emission is dominated by stellar sources, the fainter 
ones merging into an unresolved background. At 5.8 and 8 $\micron$ the stellar 
sources are fainter, and large--scale diffuse emission from the ISM of the 
Galaxy's central molecular zone becomes prominent. 
The survey reveals that 
the 8 to 5.8 $\micron$ color of the ISM emission is highly uniform across
the surveyed region. This uniform color is 
consistent with a flat extinction law and 
emission from polycyclic aromatic hydrocarbons (PAHs). 
Models indicate that this broadband color should not be expected 
to change if the incident radiation field heating the dust and PAHs 
is $< 10^4$ times that of the solar neighborhood. The few regions with unusually 
red emission are areas where the PAHs are underabundant and 
the radiation field is locally strong enough 
to heat large dust grains to produce significant 8 $\micron$ emission.
These red regions include compact H II regions, Sgr B1, and wider 
regions around the Arches and Quintuplet Clusters. In these regions the radiation field 
is $\gtrsim 10^4$ times that of the solar neighborhood.
Other regions of very red emission indicate cases where thick dust clouds 
obscure deeply embedded objects or very early stages of star formation. 
\end{abstract}

\keywords{dust, extinction --- Galaxy: center --- infrared: ISM}

%%%%%%%%%%%%%%%%%%%%%%%%%%%%%%%%%%%%%%%%%%%%%%%%%%%%
\section{Introduction}

The Galactic Center (GC) is naturally a region of great interest. 
The gravitational center of the Galaxy is marked by the supermassive
black hole at the radio source Sgr A*. 
Remarkably, two young, dense  stellar disks have been discovered 
within 0.5 pc orbiting Sgr A* (Paumard et al. 2006).
Further out is a molecular ring or the ``circumnuclear disk'' enclosing 
an ionized a mini--spiral structure (e.g. Christopher et al. 2005). 
These features are within a radius of $\sim7$ pc. 
A concentration of molecular material is observed out to a radius of $\sim200$
pc along the Galactic Plane; this region is known as the Central Molecular
Zone (CMZ). Aside from the stellar bulge of the Galaxy, the CMZ encompasses 
nearly all of the Galactic structures and objects that are 
characteristic of, or unique to, the GC.

The CMZ is estimated to contain about 10\% of the Galaxy's molecular gas 
content (e.g. Morris \& Serabyn 1996; Mezger 1978), or about 
$5-10\times10^7$ M$_{\sun}$ (e.g. Armstrong \& Barrett 1985; Stark et al. 1989). 
It has a two-component structure. The inner region 
consists of a set of low velocity clouds lying on or close to 
the Galactic plane, including the giant Sagittarius molecular 
clouds, which together comprise the ``disk population'' (Morris 
\& Serabyn 1996). The outer set of clouds form a comparatively 
high velocity (130--200 km s$^{-1}$) quasi--continuous ring structure 
that surrounds the first one at a radius of $\sim180$ pc (the 
so-called ``180-pc molecular ring''; Morris \& Serabyn 1996). 
Analysis of NH$_3$ and H$_3^+$ absorption lines in the CMZ has 
indicated that the molecular gas contains a widespread, relatively hot component
($T \gtrsim 175K$, Wilson et al. 1982; Oka et al. 2005), 
which is not found apart from dense molecular cores elsewhere in the Galaxy.
Analysis by H\"uttemeister et al. (1993) indicates
the molecular gas has at least two phases: hot phase 
($T \sim 200$K, $n_{H_2} \lesssim 10^3$ cm$^{-3}$) with a filling factor
of $\sim 25\%$, and a cooler component 
($T \sim 25$K, $n_{H_2} \gtrsim 10^4$ cm$^{-3}$) which fills the remaining
75\% of the volume of the molecular clouds.

The average dust temperature in the CMZ appears to be much lower than 
that of the gas, $T_{dust} \sim 25-30$ K (Odenwald \& Fazio 
1984; Cox \& Laureijs 1989).
The much higher gas temperatures 
therefore indicate the presence of some efficient gas 
heating mechanisms, and supersonic winds and turbulence 
had been suspected; the large gas line widths, 15--50 km s$^{-1}$, 
are consistent with such ideas, although the details of 
the mechanisms are debated (SNRs, outflows, viscous 
heating effects, etc.).
Observations of widespread SiO emission indicates the recent or current 
presence of strong shocks in the CMZ (Martin-Pintado et al. 1997; 
H\"uttemeister et al. 1998).
The discrepancy between dust and gas temperatures has recently
been accounted for in terms of the impact of cosmic ray electrons 
heating molecular clouds directly and raising their 
temperatures (Yusef-Zadeh et al. 2007b).

In addition to the molecular gas, the CMZ is also infused with 
neutral and ionized components as revealed by enhanced infrared continuum and 
line emission (e.g.  Rodr\'iguez-Fern\'andez et al. 2004, 2005),
and by enhanced radio continuum emission (e.g. Altenhoff et al. 1979).

The launch of the {\it Spitzer Space Telescope} (Werner et al. 2004)
provides a new opportunity to investigate the GC at mid-- to far--infrared 
wavelengths with unprecedented sensitivity and spatial resolution.
{\it Spitzer's} Infrared Array Camera, IRAC, (Fazio et al. 2004) 
provides $5'$ field of view imaging capabilities with $<2''$ spatial resolution
at 4 broadband mid--infrared channels nominally at 3.6, 4.5, 5.8, and 8 $\micron$.
We used IRAC to carry out a $2\arcdeg \times1\fdg4$ 
($\sim 280\times 200$ pc) survey of the GC region, which includes the 
entire CMZ. 
An initial summary of the results of this survey 
has been presented by Stolovy et al. (2006). A more extensive overview will be 
provided by S. Stolovy et al. (2008, in preparation). 
The measurement, cataloging, and characterization of point 
sources observed in our IRAC GC survey are described by Ram\'irez et al. (2008). 
In IRAC's 3.6 and 4.5 $\micron$ channels, the GC is dominated by stellar 
emission, both for resolved sources and in the unresolved background. 
At 5.8 and 8 $\micron$ wavelengths, most point
sources become significantly fainter, and a strong, structured background 
produced by the diffuse ISM is dominant. 
In this study, we focus on the diffuse ISM as revealed 
by IRAC. In particular we investigate what the broadband colors 
indicate about the nature of the dust producing the emission,
and the environment in which that dust is found.

Section 2 of this paper contains a brief summary of the IRAC observations.
Section 3 presents the colors observed for the diffuse emission
and for a few, more compact sources of special interest. Mostly 
this focuses on the 5.8 to 8 $\micron$ color of the ISM emission. 
The shorter wavelength colors are dominated by starlight and extinction
(reddening) effects. Section 4 discusses the implications of the 
uniformity of the color on large scales and the few exceptional 
regions on small scales. Though limited to the single 
5.8 to 8 $\micron$ color, our analysis yields several general robust 
conclusions concerning the relative abundance of PAHs, 
the strength of the interstellar radiation field, and the 
shape of the extinction law in the GC. The last section (\S5)
contains a summary of this paper.

%%%%%%%%%%%%%%%%%%%%%%%%%%%%%%%%%%%%%%%%%%%%%%%%%%%%
\section{Observations}
We observed the central region of the Galaxy using 12 Astronomical 
Observation Requests (AORs) which mapped the region in 
$\sim0\fdg25$ wide strips at roughly constant declination (a restriction due to {\it Spitzer's}
pointing constraints and the low ecliptic latitude of the field). Most strips were $\sim1\fdg75$ long,
except that the one containing Sgr~A was split into two parts (to mitigate the potential effects of 
latent images, with the brightest regions of the survey scheduled last). 
Additionally, those strips at the highest and lowest declinations were 
shorter, in order to limit the area of complete coverage to $|l| < 1\fdg0$ and $|b| < 0\fdg7$. 
The total area covered in all four channels was $\sim3.5$ square degrees, 
including the additional irregular edges of the region.
The medium scale 5--point Gaussian dither pattern was used. This pattern makes small offsets 
to the telescope pointing with an average offset of $\sim30''$. We did not use
the smaller scale pattern because it would have provided insufficient 
dithering for mitigating the stray light 
artifacts of bright stars, and would have hampered self--calibration of the 
data (see the Appendix). 
Frame times of 2 seconds were used throughout the survey. No regions of extended
emission were saturated except at Sgr A at 8 $\micron$, but more than $10^4$ bright point sources 
were saturated in at least one channel (Ram\'irez et al. 2008). 
Most saturated sources are isolated point sources, but 
high densities of saturated sources are found within the bright cluster at Sgr A and the Quintuplet Cluster.
These areas, and a sample of 12 other saturated sources, were re-observed in subarray mode with  
0.02--second frame times to obtain unsaturated data. All analysis in this work was performed on 
mosaicked images that incorporated the subarray data for these sources.

We constructed our own mosaics from the total set of Basic Calibrated Data (BCD, the standard 
{\it Spitzer} data products intended for detailed scientific analysis) from all AORs.
The primary description of our processing is given by 
Stolovy et al. (2006, 2008 in preparation). In the appendix of the 
present paper, we provide additional detail
concerning aspects of the processing that are particularly relevant in 
studying the brightness and color of extended emission. 

%%%%%%%%%%%%%%%%%%%%%%%%%%%%%%%%%%%%%%%%%%%%%%%%%%%%
\section{Analysis}

\subsection{The Diffuse ISM}
\subsubsection{The Large-Scale Distribution of 8 $\micron$ Intensity and 8/5.8 $\micron$ Color}

\begin{figure}
\includegraphics[scale=0.80,angle=180]{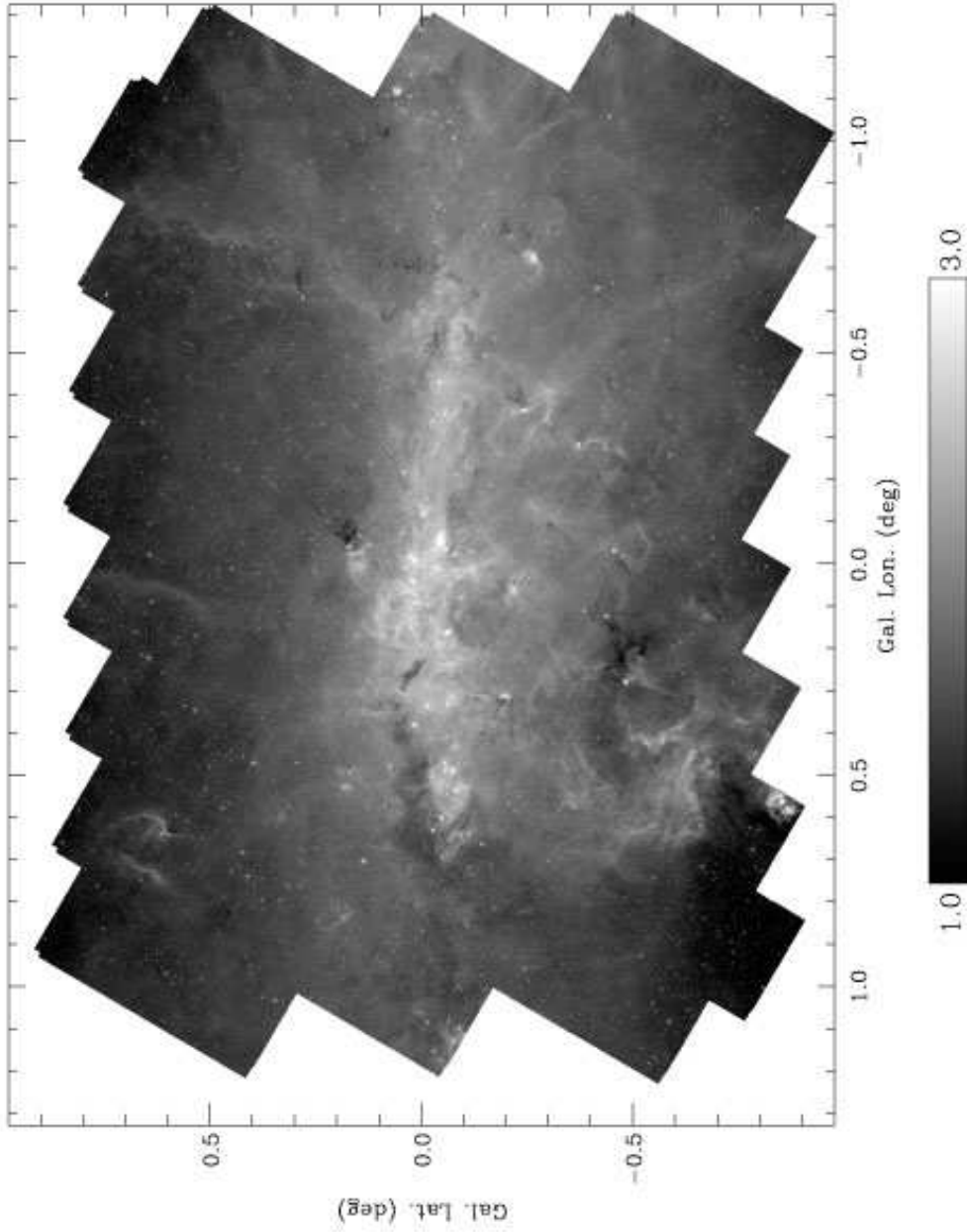}
\caption{The mosaic of the {\it Spitzer} IRAC Galactic Center survey at 8 $\micron$. 
Intensities are scaled logarithmically from 10 to 10$^3$ MJy sr$^{-1}$.
The brightest diffuse emission is correlated with the Central Molecular Zone (CMZ).
\label{fig_ch4}}
\end{figure}

\begin{figure}
\plotone{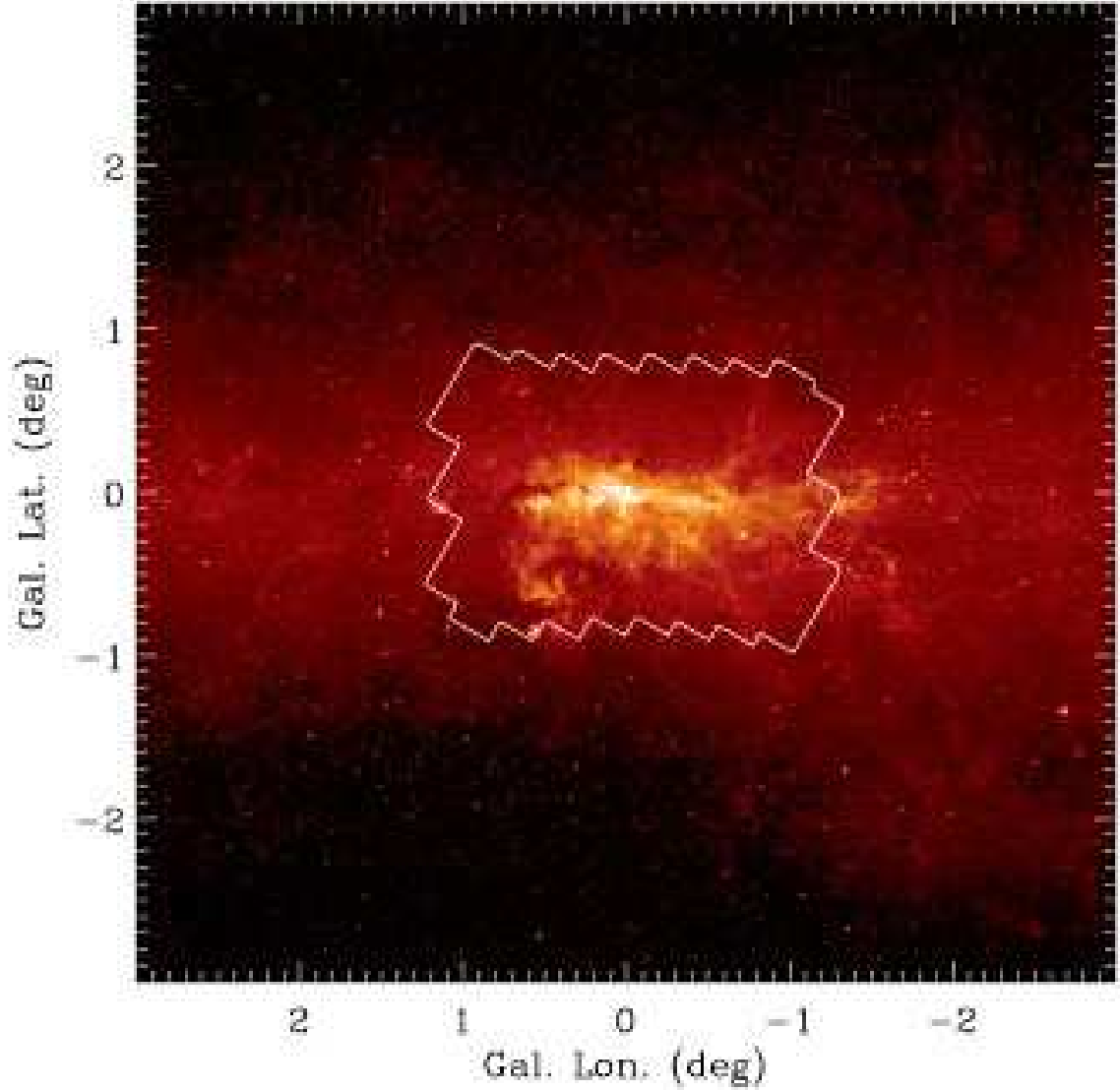}
\caption{Galactic Center region at 8.28 $\micron$ as seen by the MSX
mission (logarithmic intensity scale). The white contour indicated the 
extent of our Spitzer IRAC survey. The bright Central Molecular 
Zone at a galactocentric radius of $r < 100$ pc is clearly distinct 
from the wider span of the inner galaxy $3\arcdeg > l > -3\arcdeg$
($r < 400$ pc).
\label{fig_msx}}
\end{figure}

\begin{figure}
\plotone{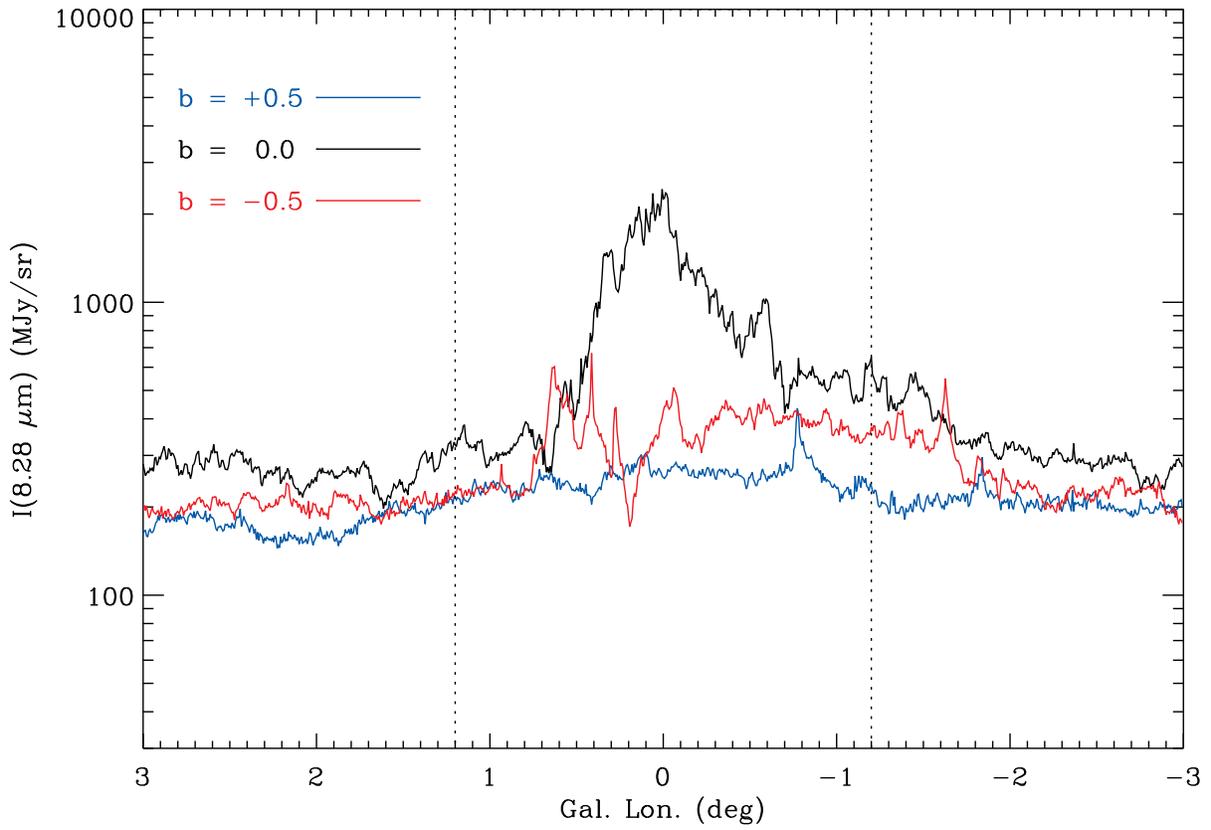}
\caption{Intensity slices through the MSX 8.28 $\micron$ data at several latitudes.
The data have been binned to $24"$ pixels and medians are taken over a $6\farcm8$
swath in latitude. The slice at $b = 0$ shows that the structures of the 
Central Molecular Zone are 2 -- 10 times brighter than the emission in 
other parts of the Galactic plane.\label{fig_msxslice}}
\end{figure}

The 8 $\micron$ mosaic of the Galactic Center region is shown in Figure \ref{fig_ch4}. 
The general structure of the ISM emission in this region appears similar to other regions 
in the Galactic plane as revealed by the GLIMPSE survey (Benjamin et al. 2003).
Both in the GC and other regions, much of the ISM emission is resolved into
elongated arcs and filamentary structures (S. Stolovy et al. 2008, in preparation). 
Embedded among these are very bright compact regions
that contain specific local heating sources. Superimposed infrared dark clouds (IRDCs) have a 
similar structure to the emission, but are not well-correlated with the emission.
The ISM emission has a strong unresolved background component, which is often 
suppressed in depictions meant to highlight the structure of the emission 
(such as Fig. \ref{fig_ch4}).
The most fundamental difference between emission in the GC and that in other regions
of the Galactic plane, is that the GC emission is several times brighter. 
Figure \ref{fig_msx} shows a wider view of the Galactic 
Center region as seen by SPIRIT III on the MSX mission (Price et al. 2001),
revealing that the emission within our survey region is brighter than that 
typically found in the inner galaxy (at $|l| > 1\fdg5$). 
Figure \ref{fig_msxslice} provides a 
more quantitative illustration of the relative brightness of our survey region. 
The intensity enhancement is loosely correlated with 
the large--scale distribution of molecular emission in the galactic
center region (e.g. Martin et al. 2004). 
This indicates that much of the low--latitude ($|b| < 0\fdg2$)
structure in the galactic center mosaic really is in the central 
$\sim150$ pc radius of the Galaxy, but that the emission does not directly arise 
from the molecular component of the CMZ.   

In contrast, emission at higher 
latitudes ($|b| > 0\fdg2$) often appears to be more local features
superimposed along the line of sight. One indication of this is association 
with H$\alpha$ emission, which would be too heavily extincted to be 
visible at the distance to the GC, with the possible exception of 
objects seen at the extreme latitude limits of our survey 
(Stolovy et al. 2006, 2008 in preparation). Another is the association with 
very dark clouds, which must lie relatively close to the sun. Otherwise,
the general foreground emission along the line of sight would reduce the 
contrast of these features. The large angular scale and 
presence of fine detail in the higher latitude features also 
suggest they are relatively nearby clouds along the lines of sight.
Pauls \& Mezger (1975) use the velocity and line widths of 6 cm radio recombination 
lines to distinguish emission from the GC (large velocities and/or broad lines), 
from more local emission along the line of sight (velocities near zero, and narrow
line widths). All of the regions they examined at $|b| > 0\fdg16$ are
characterized as being along the line of sight, rather than at the GC.

\begin{figure}
\includegraphics[scale=0.80,angle=180]{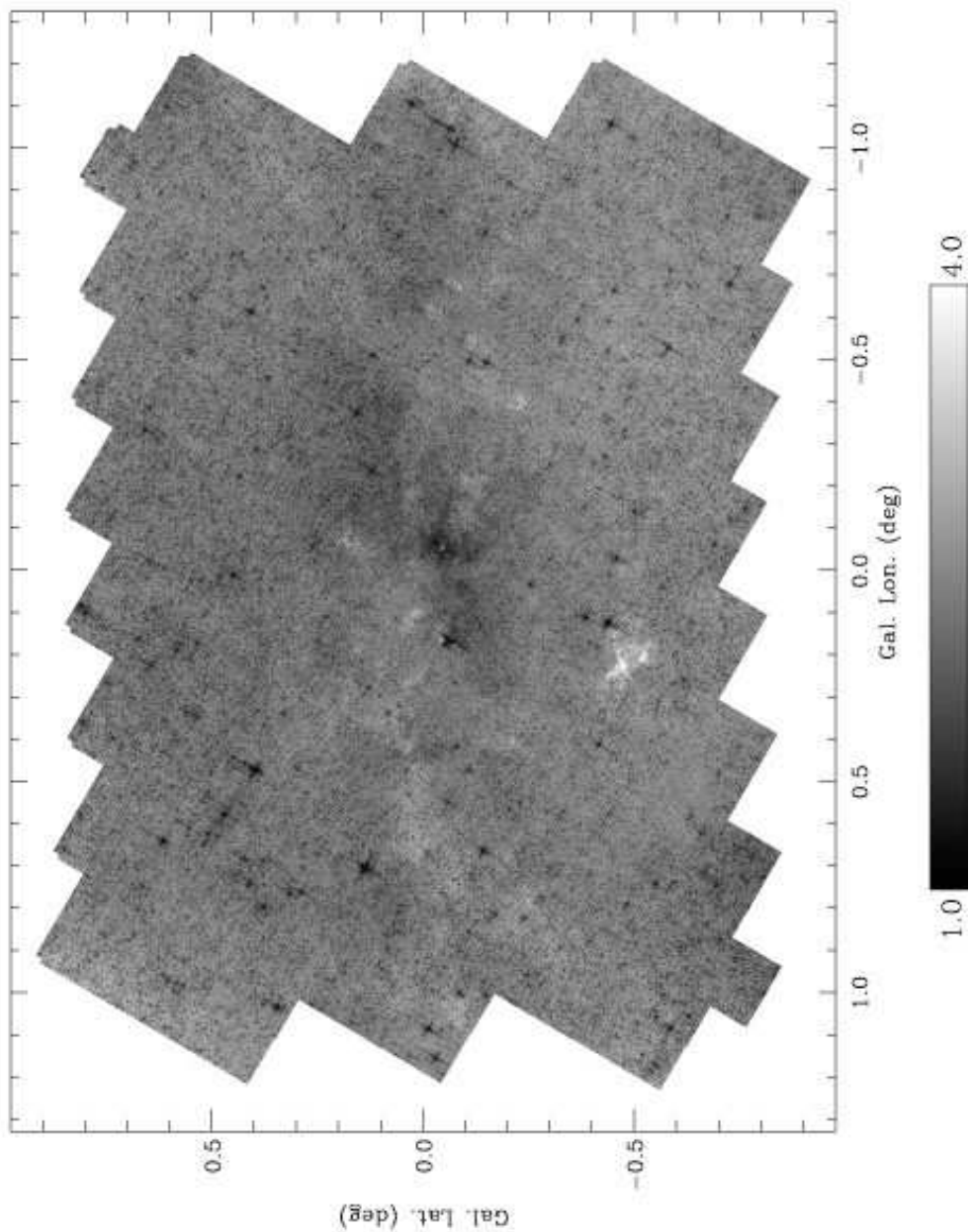}
\caption{Galactic Center region depicted in the $I (8 \micron) / I(5.8 \micron)$ color
(linear grayscale).
This color shows little variation or correlation with the 8 $\micron$ intensity 
(Fig. \ref{fig_ch4}). Stellar sources have $I (8 \micron) / I(5.8 \micron) < 1$ and 
appear black here. Only a few extended regions have redder than average colors. Several 
of these are shown in detail in Figs. \ref{red_regions} and \ref{red_regions2}. 
The distinct complex near $(l,b) = (0\fdg24,-0\fdg45)$ 
is at the darkest of the IRDCs, and thus may be affected by small 
errors in subtraction of the zodiacal light emission or detector offsets. 
\label{fig_color34}}
\end{figure}

Figure \ref{fig_color34} shows the ratio of 8 $\micron$ / 5.8 $\micron$ emission,
i.e. the $I(8 \micron) / I(5.8 \micron)$ color.
In this ratio virtually all of the diffuse emission features 
disappear, indicating that they have a very uniform color that 
is independent of brightness. Dark regions in the ratio map indicate
bluer colors associated with numerous individually resolved stars,
and with the unresolved stellar background which becomes comparatively 
bright in the innermost regions of the galaxy (central $\sim10'$). 
There are only a few unusually red point sources. Two of these are
asteroids (1567 Alikoski and 459 Signe), and two others are PN M1--26 (HD 316248; 
Sahai \& Trauger 1998) and the candidate PN 359.491+0.316 (IRAS 17399--2910; 
Becker et al. 1994). Brighter (i.e. redder) extended regions
in the ratio map are associated with a few special types of emission features 
which will be discussed below, and with the darkest of the IRDCs
[e.g. the complex near $(l,b) = (0\fdg24, -0\fdg45)$].
The apparent redness of the IRDCs may be an artifact of 
residual errors in the subtraction of the zodiacal light or 
instrumental offsets (see the Appendix and Table 1). 
Small offset errors would have the greatest 
impact on the faintest regions of the images, which are the IRDCs
and regions at the extreme latitudes along the edges of the mosaics.

\subsubsection{ISM Colors Corrected for the Effect of Unresolved Starlight}
To better isolate the ISM emission we have subtracted stellar emission 
in a manner similar to that applied to external galaxies (Pahre et al. 2004).
In effect, we decompose the 8 and 5.8 $\micron$ emission into ISM and stellar 
components using the 4.5 $\micron$ emission as strong tracer of the starlight. 
We assume that there exist fixed spatial templates of each component, such that 
\begin{equation}
I(\lambda) = \alpha_{\lambda}I_{ISM} + \beta_{\lambda}I_{stellar}
\end{equation}
where $I(\lambda)$ is the observed emission at $\lambda = 4.5$, 5.8 or 8 $\micron$,
$I_{ISM}$ and $I_{stellar}$ are fixed (but unknown) templates of the ISM and stellar 
emission, and $\alpha_{\lambda}$ and $\beta_{\lambda}$ are the scaling factors
for the respective templates. The ratios of the $\alpha_{\lambda}$ or 
$\beta_{\lambda}$ coefficients yield the colors of the ISM and stellar emission 
at the relevant wavelengths.
If the ISM emission were negligible at 4.5 $\micron$ and stellar emission 
were negligible at 8 $\micron$, then $\alpha_{4.5 \micron} = \beta_{8 \micron} = 0$
and a linear least squares fit of the form 
\begin{equation}
I(5.8 \micron) = A\ I(8 \micron) + B\ I (4.5 \micron) + C
\end{equation}
could be used to determine: $A = \alpha_{5.8\micron}/\alpha_{8\micron}$, 
the mean $5.8 / 8 \micron$ color of the ISM; 
$B = \beta_{5.8\micron}/\beta_{4.5\micron}$, 
the mean $5.8 / 4.5 \micron$ color of the starlight; 
and $C$, a constant term representing 
a linear combination of the mean systematic errors (including those of the detector offsets, 
zodiacal light, and the stellar model) in all three channels. 
However, because there is no {\it a priori} knowledge of the magnitude or sign
of the systematic errors in each channel, we attribute this term to only 
the 5.8 $\micron$ channel, which exhibits the greatest instability in detector offset.

In reality, the longer wavelength stellar emission and shorter 
wavelength ISM emission are not negligible. Therefore 
the interpretation of the $A$, $B$, and $C$ coefficients 
derived in Eq. 2 is slightly more complicated.
To derive both $\beta_{5.8\micron}/\beta_{4.5\micron}$ and 
$\beta_{8\micron}/\beta_{4.5\micron}$ from $A$ and $B$, 
we must assume the color of the starlight at one pair of wavelengths. 
We choose to set the same $[5.8]-[8.0] = +0.04$ mag 
theoretical color that was adopted by Pahre et al (2004), as these colors should 
be least affected by extinction. Then we can appropriately scale and subtract
$I (4.5 \micron)$ from the 5.8 and 8 $\micron$ emission to remove the mean stellar
emission at each of these wavelength. 
The derived $[4.5]-[5.8]$ and $[4.5]-[8.0]$ colors are about 0.1 mag redder than those
used by Pahre et al. (2004). 
Because $I (4.5 \micron)$ also contains a
weak $I_{ISM}$ component, after this subtraction of the starlight the ISM 
emission is slightly underestimated. The factor by which the ISM is reduced
can be calculated from $A$ and $B$, if we again assume the color of the ISM 
emission at one set of wavelengths. We use the ``BARE-GR-FG'' 
interstellar dust model of Zubko, Dwek, \& Arendt (2004) to calculate that
$\alpha_{4.5 \micron}/\alpha_{5.8 \micron} = 0.194$. From this we find that using 
$I (4.5 \micron)$ as a template of the starlight leads to underestimating the 
ISM emission by factors of 0.85 and 0.97 at 5.8 and 8 $\micron$ respectively. 
To summarize we construct:
\begin{equation}
I_{ISM}(8 \micron) = \left[I(8 \micron) - 0.449\ I(4.5 \micron)\right] / 0.97
\end{equation}
and
\begin{equation}
I_{ISM}(5.8 \micron) = \left[I(5.8 \micron) - 0.777\ I(4.5 \micron)\right]/0.85 + 1.68.
\end{equation}

\begin{figure}
\plotone{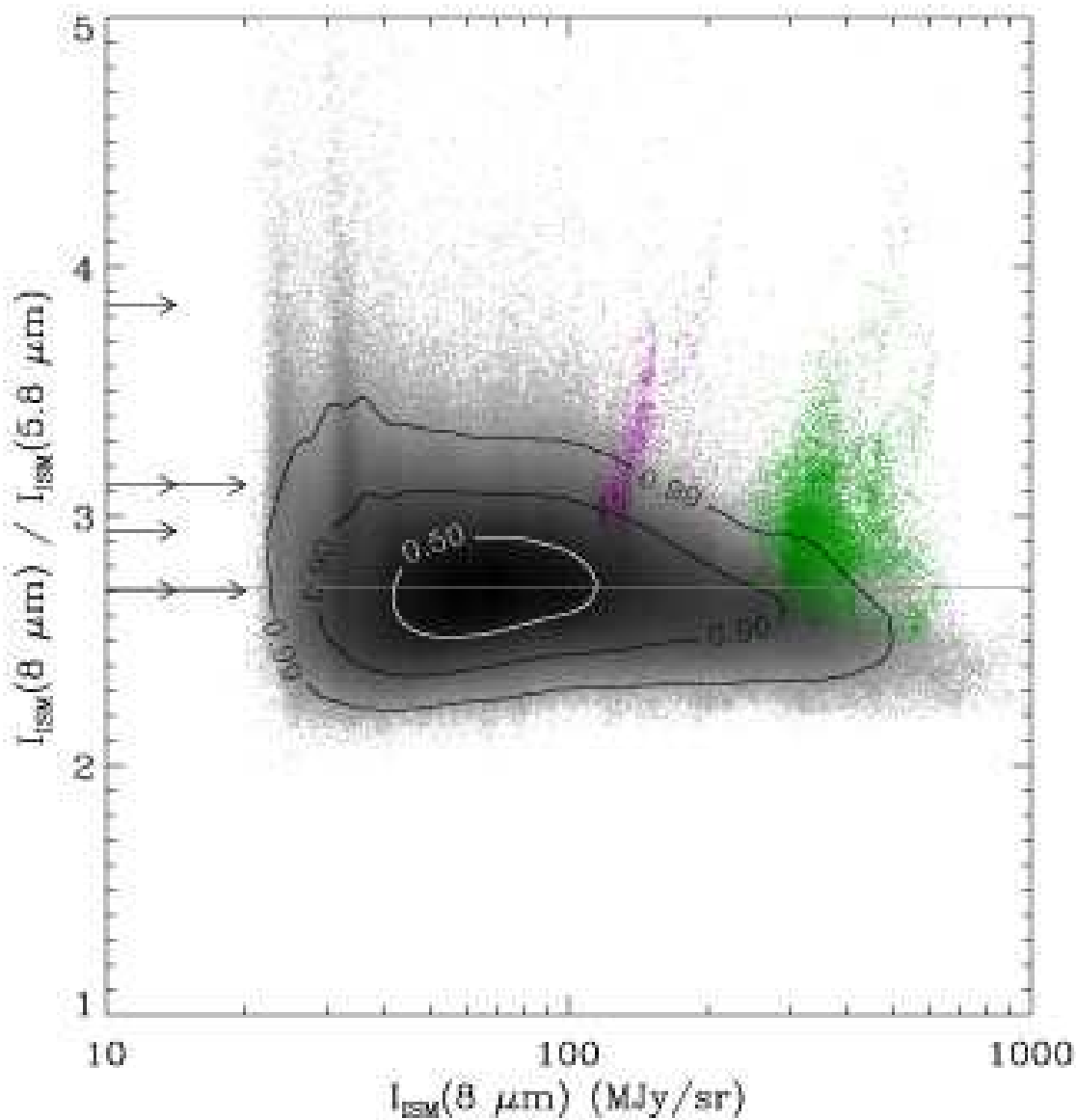}
\caption{The 8/5.8 $\micron$ color vs. brightness in the Galactic Center region.
The shading represents the diffuse emission. Contours indicate the 
regions containing 50, 90, and 99\% of the data.
The solid line shows the median color (2.71) of the diffuse emission, 
and the arrows at left indicate the colors reported for 6 other 
regions of the Galactic plane by Flagey et al. (2006). 
The cluster of points shaded in green indicate those found in the shell around
the Arches Cluster (see Fig. \ref{red_regions}), and the cluster of points shaded in magenta
indicate those found in the more compact regions at higher latitudes 
(see Fig. \ref{red_regions2}). The large dispersion in the 
color at $I_{ISM}(8\ \micron) < 35$ MJy 
sr$^{-1}$ is from faint locations at $|b| > 0.5$ where errors in the 
subtraction of stellar and zodiacal backgrounds can have large effects on the colors.
\label{fig_cmd34}}
\end{figure}

The 8/5.8 $\micron$ ISM colors found after subtraction of the starlight (and offset error) 
are slightly redder and no longer differ at regions of high extinction, but otherwise appear 
quite uniform over a wide range of brightness.  This is depicted
in Figure \ref{fig_cmd34}, where the color is plotted as a function of the 8 $\micron$ 
brightness. The colors shown here are those of the integrated emission
along the line of sight. The shading of this figure is scaled to the log of the
number of pixels with a given color -- brightness combination. This logarithmic 
scaling enhances the visibility of the outliers in the color. 
Some of the unusually red extended regions that stand out in the $I(8\micron)/I(5.8\micron)$
colors (both before and after subtraction of starlight) are highlighted and
are discussed in \S3.2.
The median color ($\pm 1\sigma$ standard deviation) 
is $I_{ISM}(8 \micron)/I_{ISM}(5.8 \micron) = 2.71 \pm 0.19$,
independent of the 8 $\micron$ brightness. 
The locations of bright stellar sources 
have been excluded from this plot with an intensity threshold cut at 70 MJy sr$^{-1}$
at 4.5 $\micron$ and by excluding regions with $I(8 \micron)/I(5.8 \micron) < 2.3$
(before the subtraction of starlight), but remaining faint resolved stars 
and systematic errors in using the 4.5 $\micron$ emission as the template
cause much of the dispersion toward redder colors at low intensities ($<$50 MJy sr$^{-1}$). 
The observed $I_{ISM}(8 \micron)/I_{ISM}(5.8 \micron)$ colors in the GC are
similar to (though slightly bluer than) 
the colors observed by Flagey et al. (2006) in 6 regions at longitudes
$l = 27\fdg5$, $105\fdg6$, and $254\fdg4$. (These latter colors are indicated by the arrows at
the left edge of Figure \ref{fig_cmd34}.) The difference is established using a two--tailed
Kolmogorov--Smirnov test, which yields a maximum deviation between the cumulative 
distribution function of $D = 0.58$. For the given number of data points, if the two 
distributions are intrinsically the same, the probability of $D$ exceeding this value
is 0.02. Therefore, this test finds only a 2\% chance that the  
Flagey et al. (2006) results are consistent with a random sampling of 
the colors that we measure near the GC. However, while this statistic indicates the colors 
are likely to be different, we cannot rule out that procedural differences or systematic 
errors are the cause of the apparent differences.

\subsubsection{Shorter Wavelength Colors}
We are unable to make an analogous comparison to 
$I(8 \micron) / I(3.6 \micron)$ and $I(8 \micron) / I(4.5 \micron)$ colors 
measured by Flagey et al. (2006) because the very bright and confusing unresolved
stellar background toward the GC overwhelms nearly all diffuse ISM emission at 
the shorter wavelengths. Point source--subtracted images (see Ram\'irez et al. 2008)
provide some aid in distinguishing non--stellar 3.6 and 4.5 $\micron$ emission,
but all such regions appear to be H II regions and PDRs associated with specific
heating sources, rather than diffuse ISM heated by the 
interstellar radiation field (ISRF).

\subsection{Unusual Emission Regions}

\begin{figure}
\centering
\begin{tabular}{ccc}
\includegraphics[width=2in]{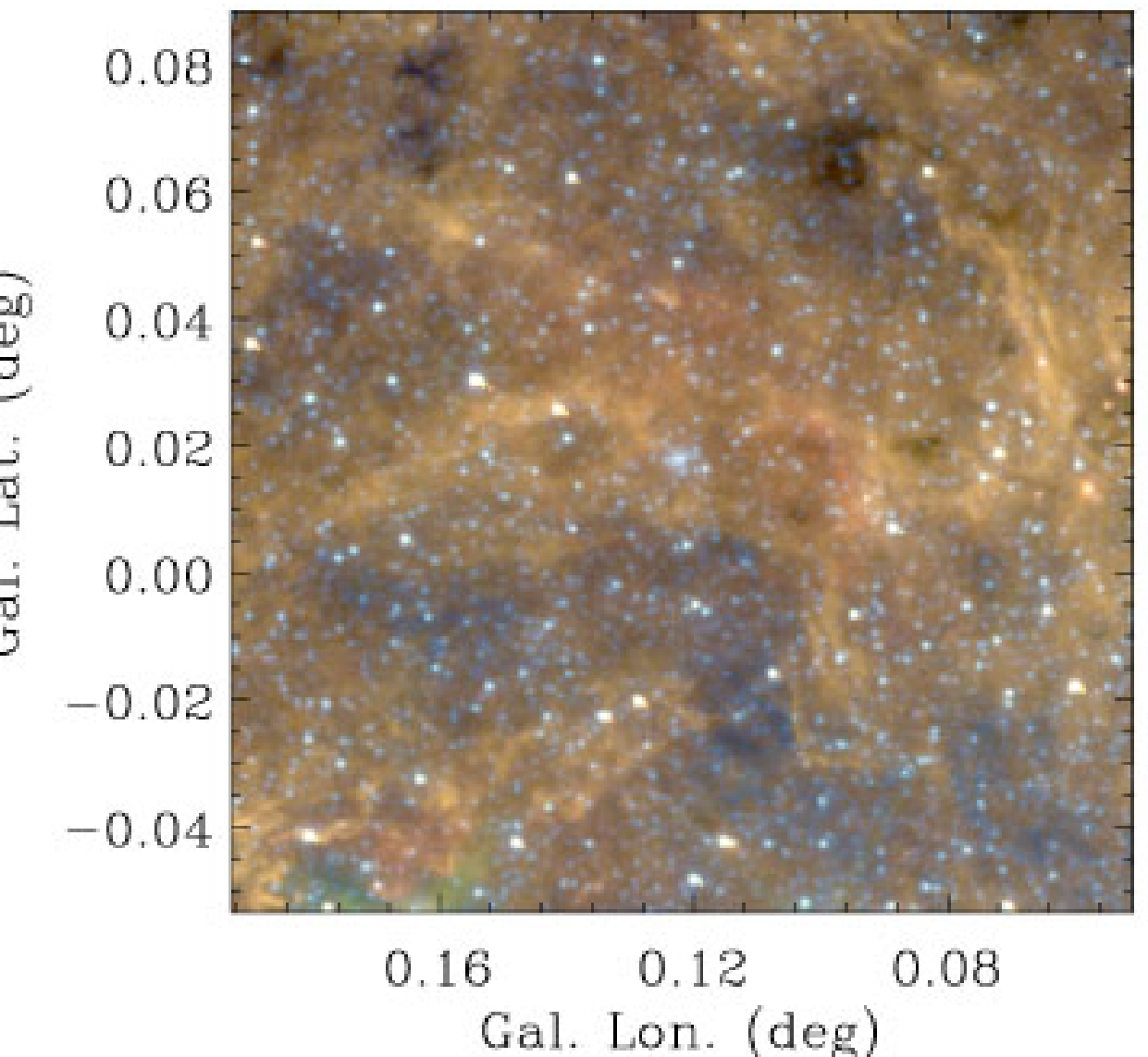}&
\includegraphics[width=2in]{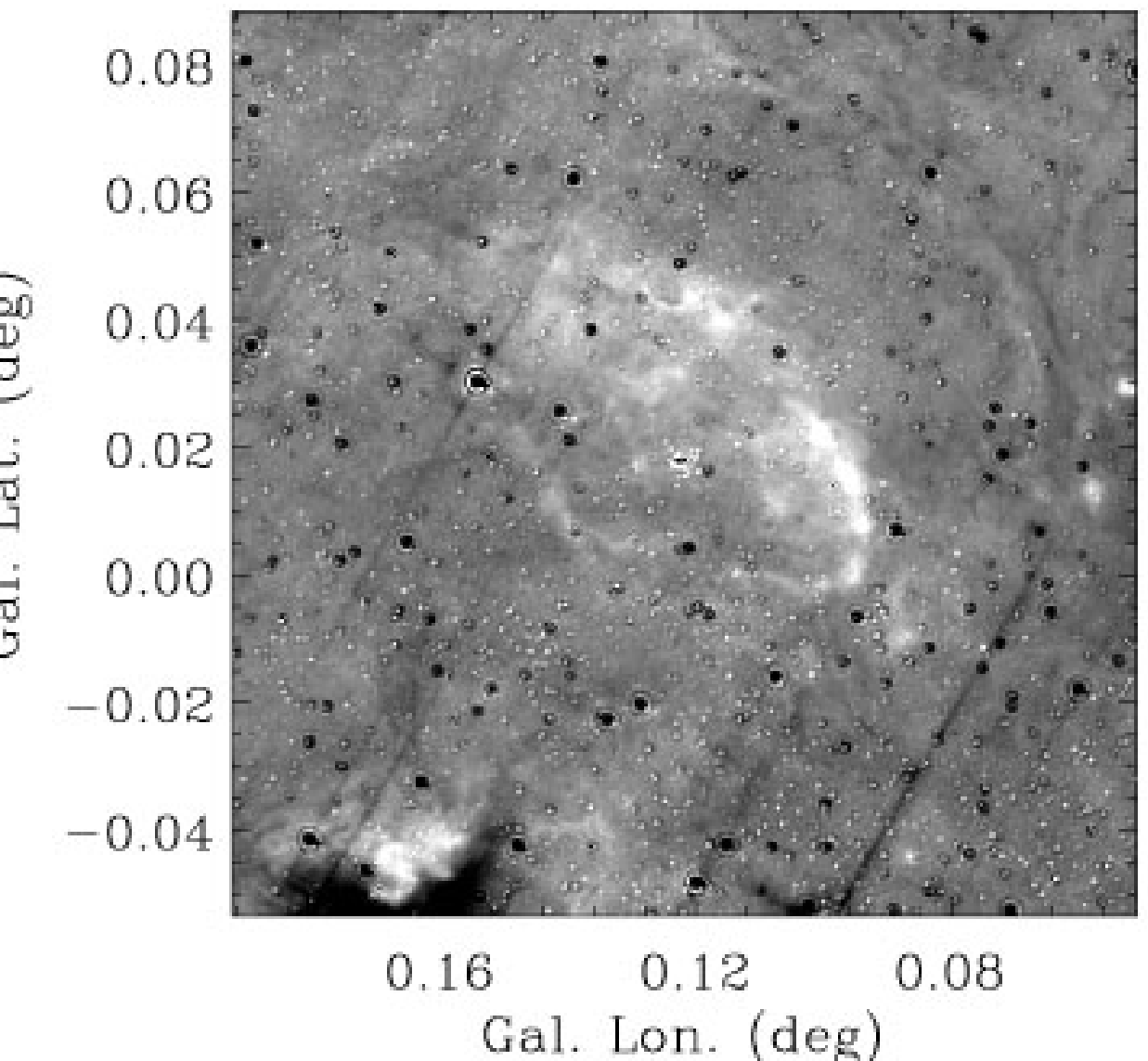}&
\includegraphics[width=2in]{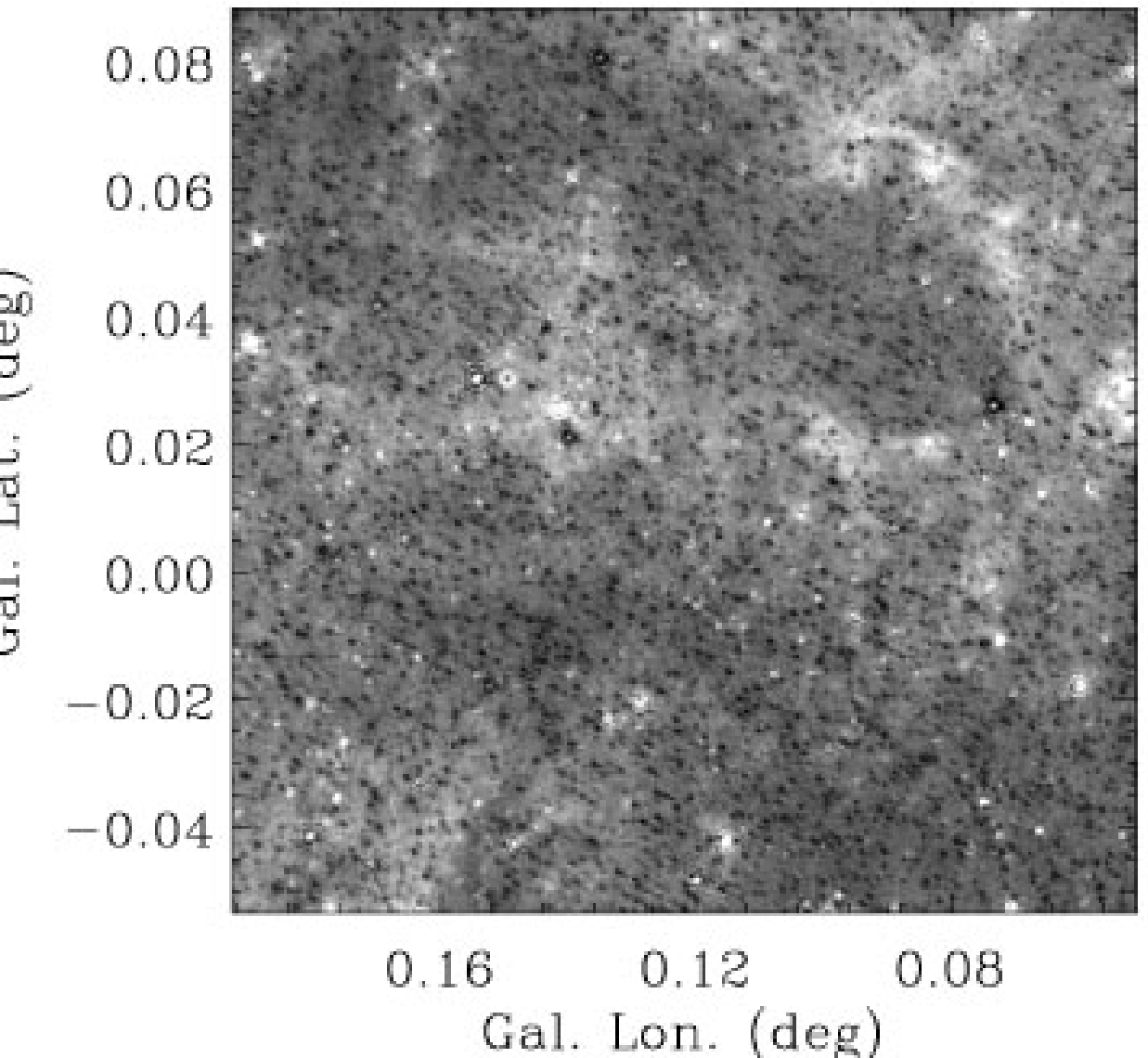}\\
\includegraphics[width=2in]{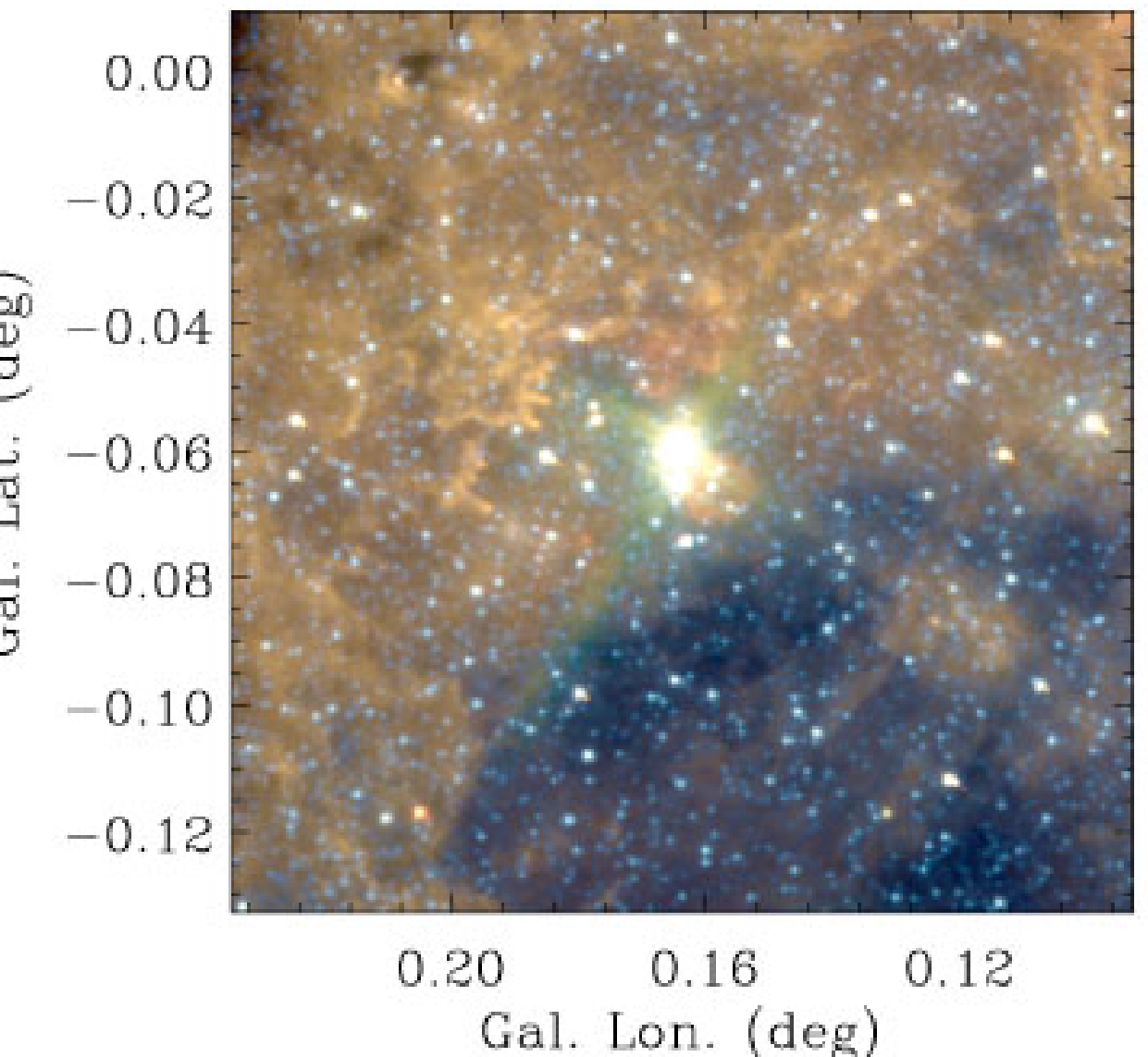}&
\includegraphics[width=2in]{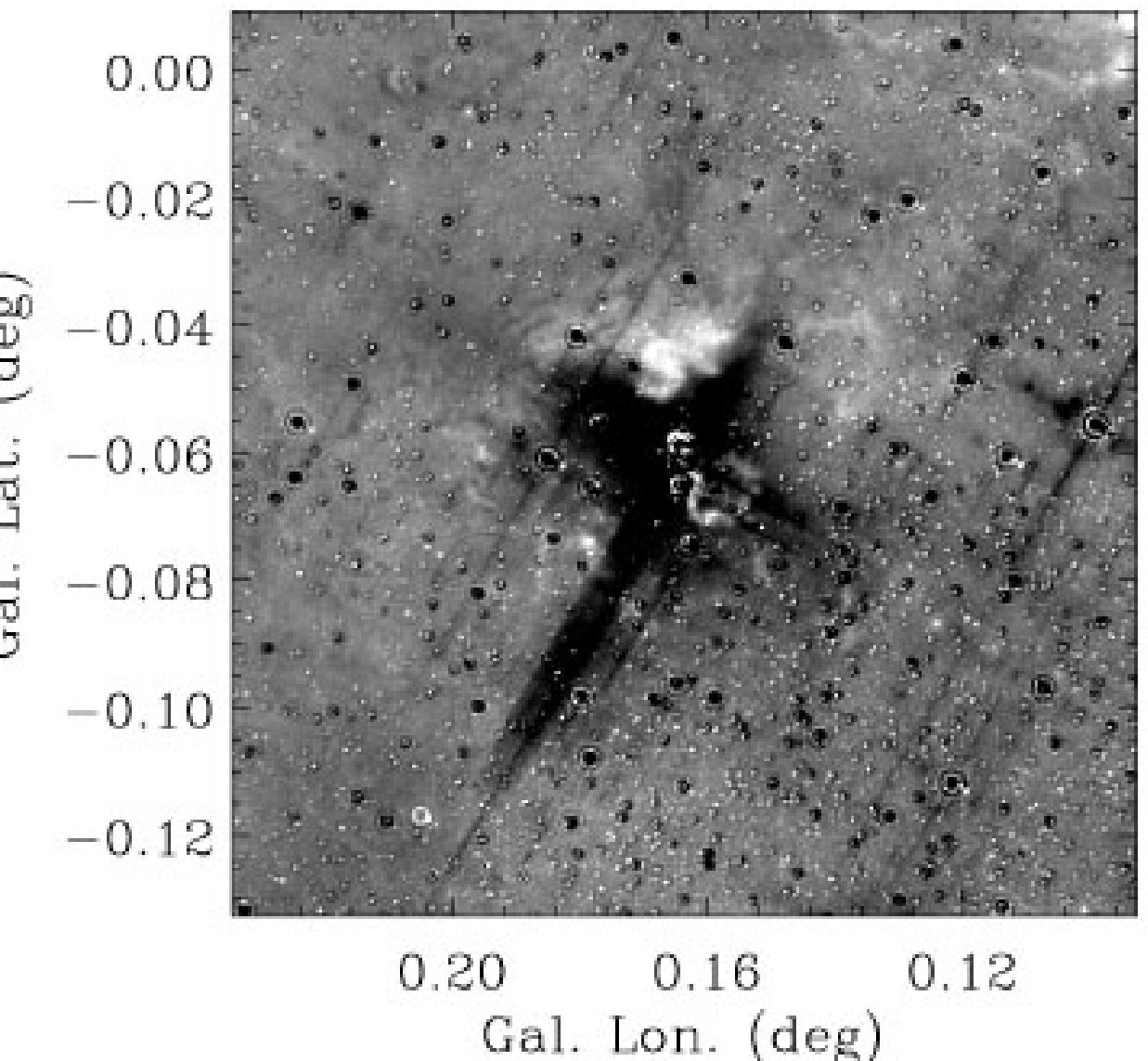}&
\includegraphics[width=2in]{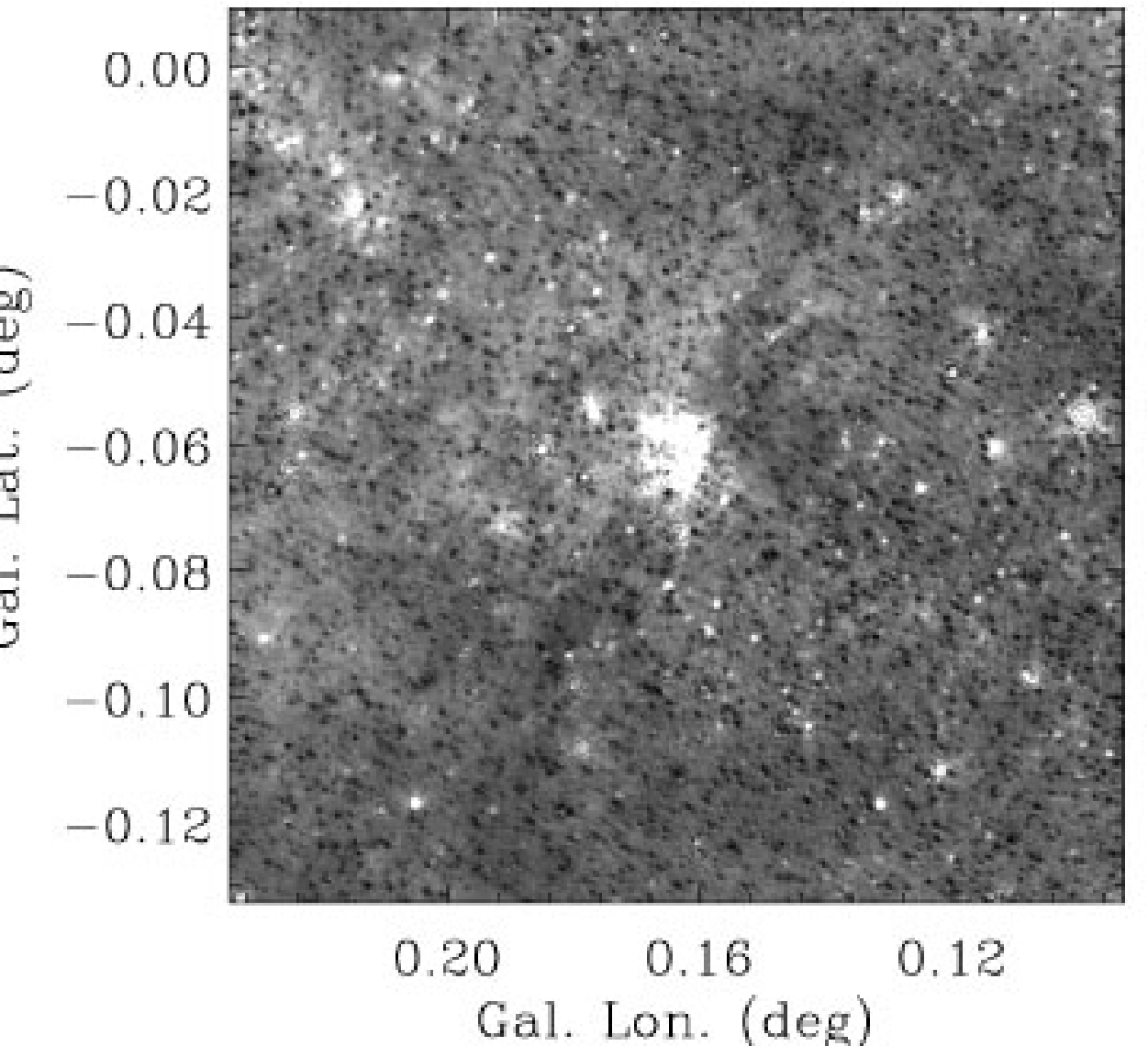}\\
\includegraphics[width=2in]{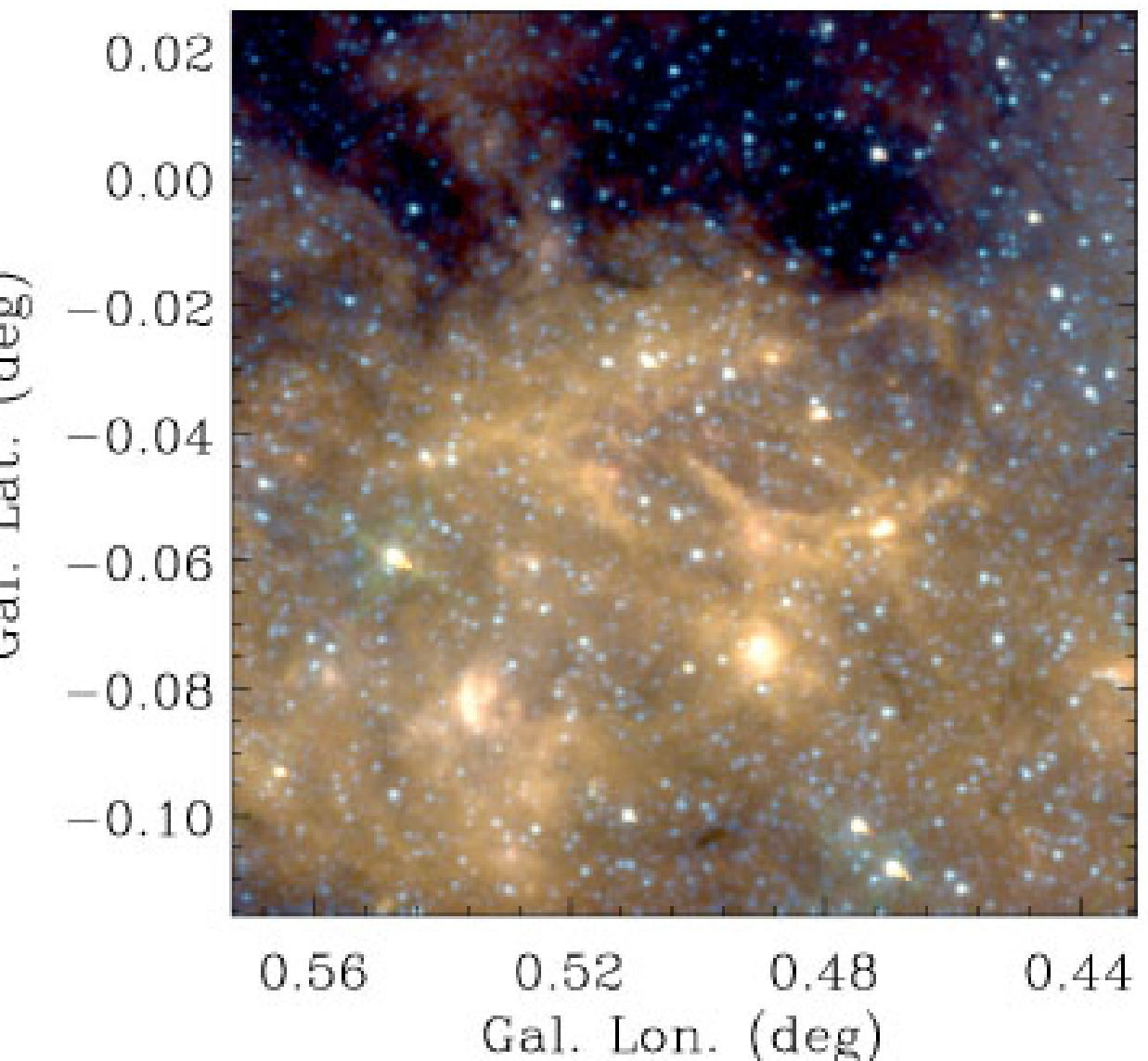}&
\includegraphics[width=2in]{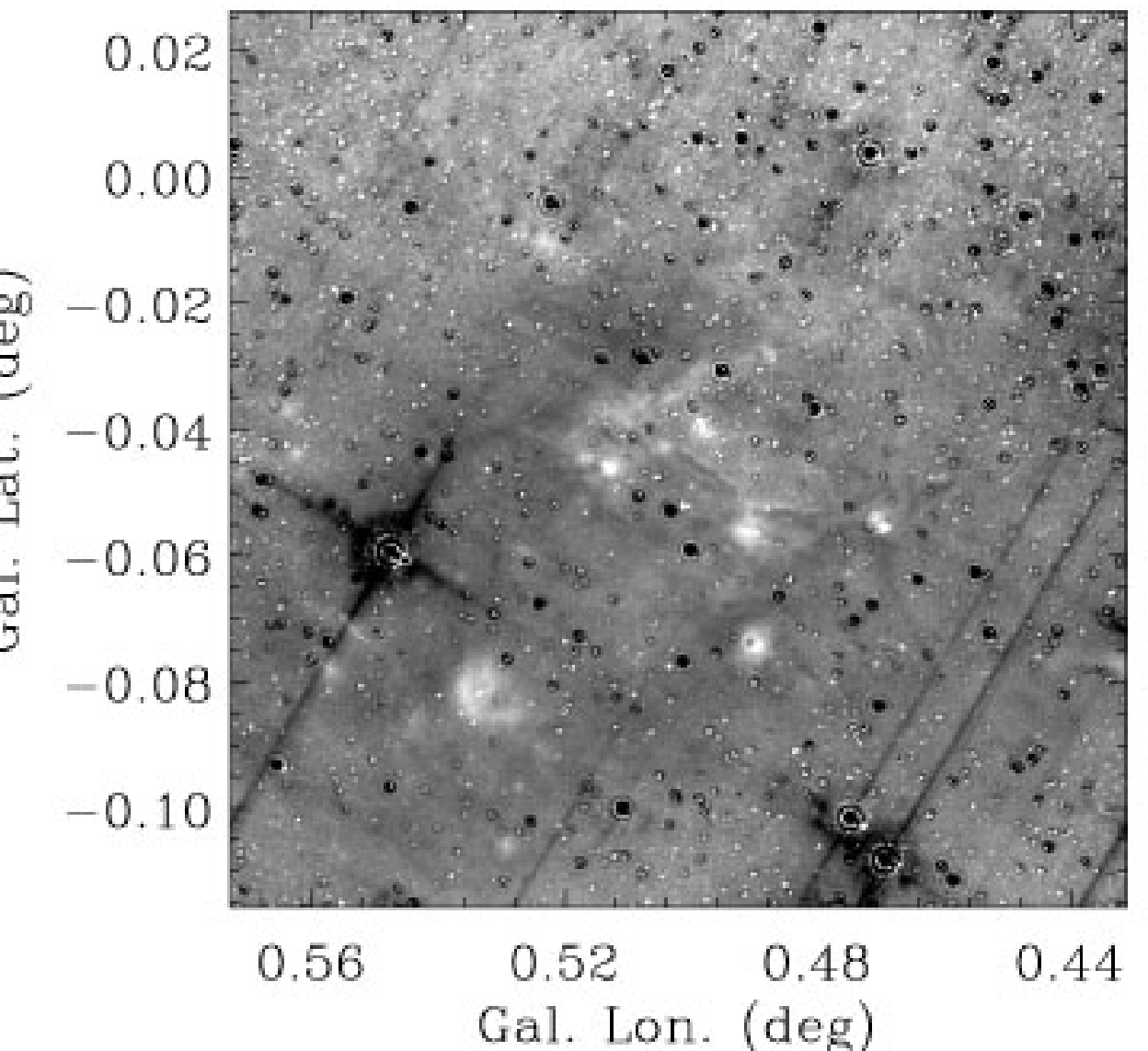}&
\includegraphics[width=2in]{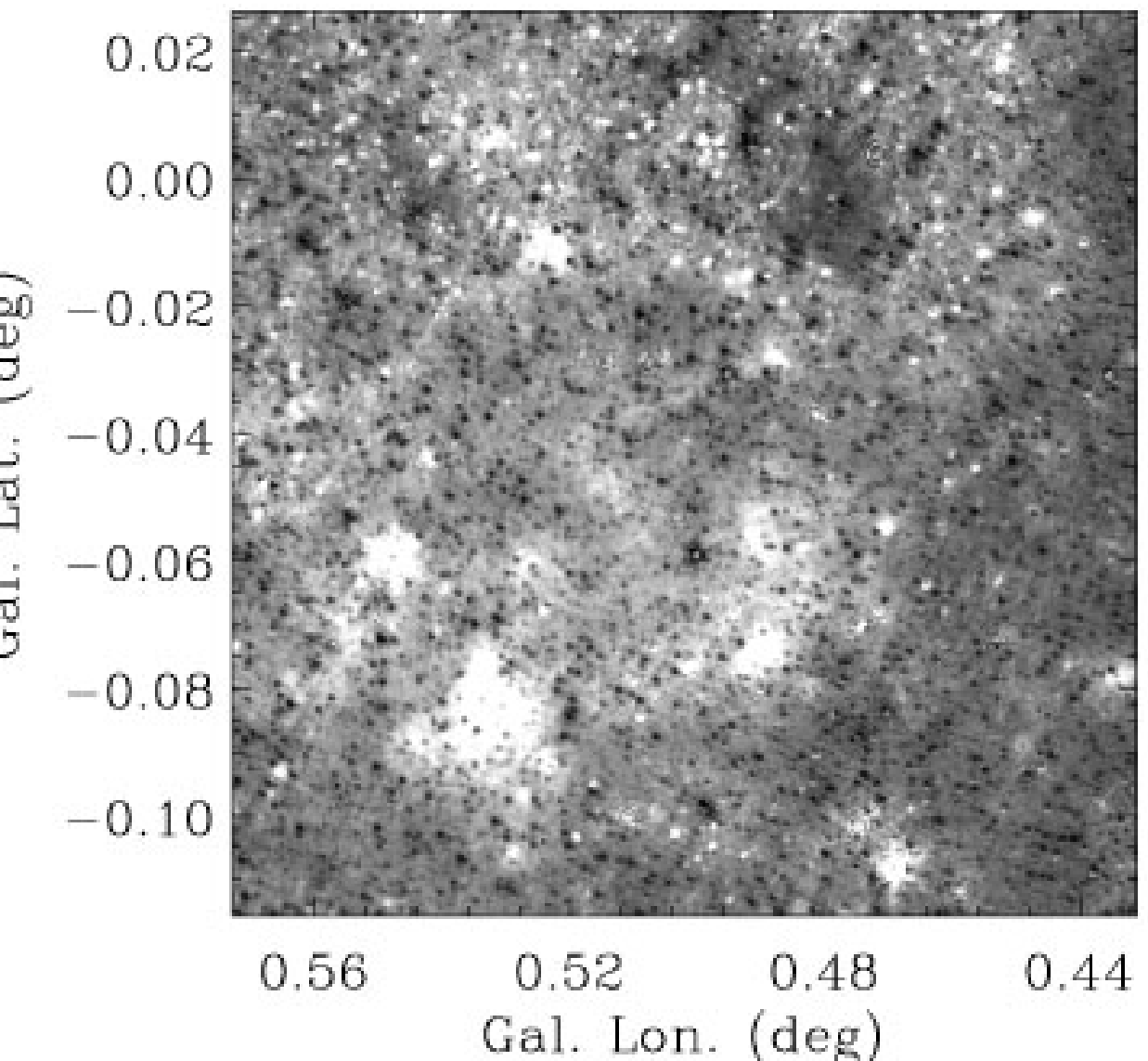}
\end{tabular}
\caption{Unusually red regions at low latitudes. The Arches Cluster region (top row),
Quintuplet Cluster region (middle row) and Sgr B1 region (bottom row). 
Left column shows 4.5, 5.8, 8 $\micron$ emission
logarithmically scaled as blue, green, and red respectively. The center column shows 
$I_{ISM}(8\micron)/I_{ISM}(5.8\micron)$ linearly scaled in the range [2.0, 4.0]. The 
right column shows $I(4.5\micron)/I(3.6\micron)$ linearly 
scaled in the range [0.5, 1.5]. 
(In the middle and right columns, lighter shades indicate redder colors.)
All these regions contains extended emission that, 
despite being relatively faint, is very red in $I_{ISM}(8\micron)/I_{ISM}(5.8\micron)$ 
colors but not in $I(4.5\micron)/I(3.6\micron)$ colors. 
\label{red_regions}} 
\end{figure}

Along the Galactic plane, there are a few large ($\gtrsim 4'$) emission 
features that stand out in the $I(8 \micron) / I(5.8 \micron)$ color map with unusually red colors.
The largest of these is an oblong shell--like structure 
centered on the Arches cluster which 
contains $\sim$200 stars with masses $>20$ M$_{\sun}$ (Figer et al. 2002). 
The shell is much more clearly delineated by the color of its emission, than 
by the 8 $\micron$ surface brightness (Figure \ref{red_regions}, top row). 
On the north, this shell seems bounded 
by the innermost of the thermal radio arches surrounding the cluster. 
The shell is only faintly discernible in the shorter wavelength colors.
A similar but smaller region appears near
the Quintuplet cluster (Figure \ref{red_regions}, middle row), which is comprised of a cluster of hot 
O/B and WR stars responsible for ionizing the nearby ISM (Cotera et al. 1996). 
However, the five cocoon 
stars (Moneti at al. 2001) that give this cluster its name 
are so bright at IRAC wavelengths that artifacts from
these sources obscure much of this excess. 

There are weaker indications
of unusually red emission in the Sgr B1 region as well (Figure \ref{red_regions}, 
bottom row). The redder infrared emission is 
associated with 5.8 and 8 $\micron$ counterparts of the ``ionized bar'' 
and ``ionized rim'' features identified in 3.6, 6 and 20 cm radio observations 
by Mehringer et al. (1992). Unlike the Arches and Quintuplet regions, however, the 
most luminous stars here are not grouped in a tight easily-identified cluster, but 
are apparently dispersed throughout the volume. 

All these extended red regions (the vicinities of the Arches 
and Quintuplet Clusters and Sgr B1)
can be distinguished in Figure \ref{fig_cmd34} as the cluster of outlier points at 
$250 < I(8 $\micron$) < 1000$ MJy sr$^{-1}$ 
and $I_{ISM}(8 \micron)/I_{ISM}(5.8 \micron) > 3.2$.
We expect that as with the compact H II regions, the
excessively red emission is caused by thermal emission from very warm dust in strong 
radiation fields. However, in these cases the excess emission appears over 
a wider region because of the larger number of luminous stars in these 
clusters and lower column densities around these more evolved systems allowing
a strong radiation field to propagate over a larger volume.

The \object{Pistol Nebula} on the edge of the Quintuplet Cluster is a different type 
of red emission region. The 8 $\micron$ IRAC data reveal an 
entire shell or bubble around the Pistol Star, with a diameter of 
$\sim22''$ (almost 1 pc at the distance of the GC). The brighter portion of this bubble 
is likely being ionized by the hot stars in Quintuplet cluster (Figer et al. 1999). 
The Pistol Nebula exhibits the most extremely red 
$I_{ISM}(8 \micron)/I_{ISM}(5.8 \micron)$ color in the 
region, as it is only marginally distinguishable at 5.8 $\micron$. 
(The 5.8 $\micron$ emission is very difficult to quantify because of strong
banding artifacts from the extremely bright stars of the Quintuplet Cluster.)
Physically, the emission from this 
region is associated with the ejecta from single star (the Pistol Star) 
which is an LBV (Figer et al. 1998). In this case, ISO 
spectroscopy of the nebula already suggests that the 8 $\micron$ excess here is due to 
a relatively red continuum (with deep silicate extinction at 10 $\micron$) and 
lack of PAH emission (Moneti et al. 1999). The [Ar II] 6.99 $\micron$ is the strongest line
between 5 and 10 $\micron$, but it it not bright enough to be the dominant cause for the 
red color. There are no other similar (bright, sharply defined, and very red) 
circumstellar bubbles evident in our galactic 
center survey. Two partial shells around massive
stars near the GC have been identified by Mauerhan et al. (2007). Neither of
these shells exhibit especially red $I_{ISM}(8 \micron)/I_{ISM}(5.8 \micron)$
color, although in one case red emission is present interior to the shell.

\begin{figure}
\centering
\begin{tabular}{ccc}
\includegraphics[width=2in]{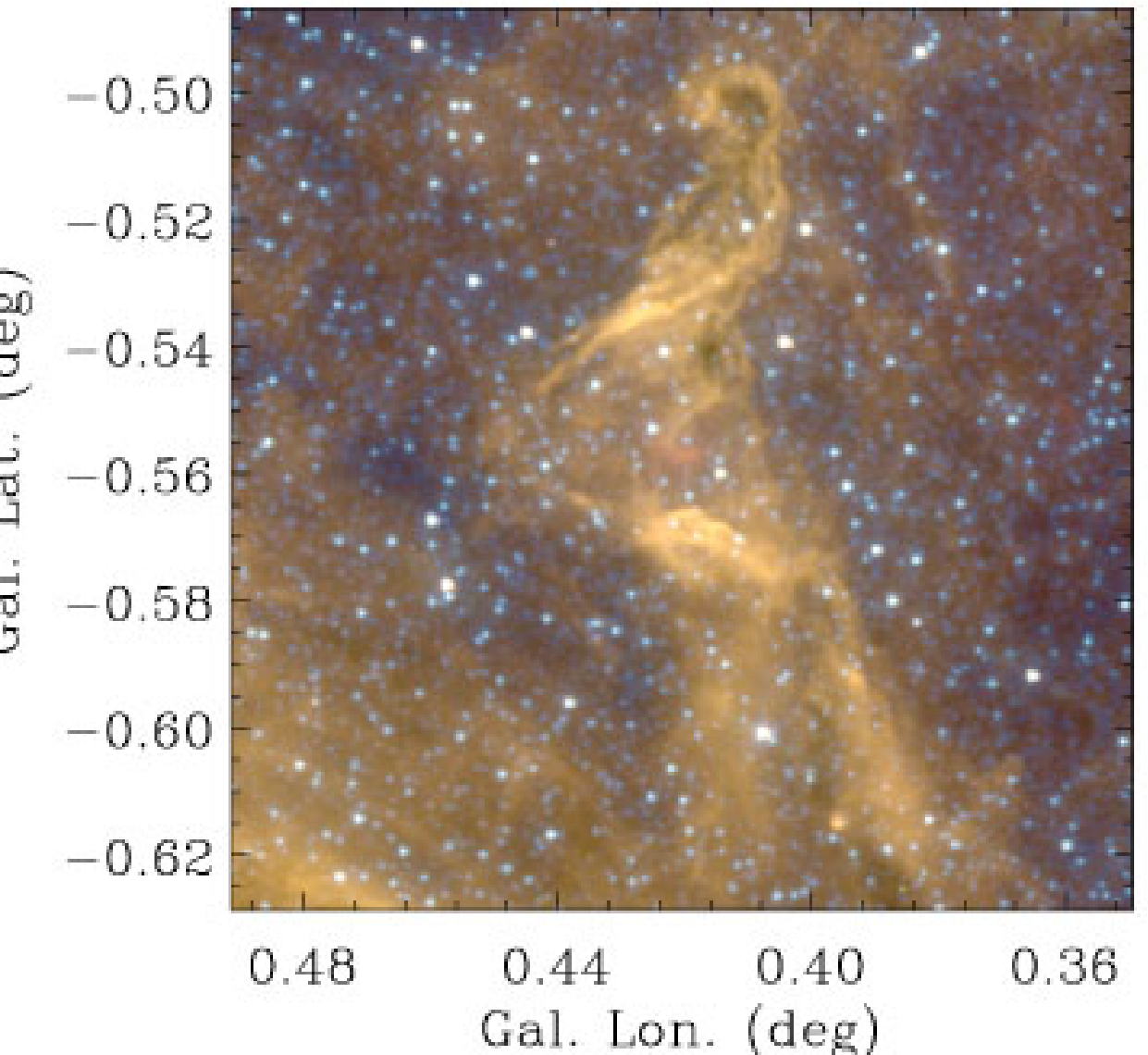}&
\includegraphics[width=2in]{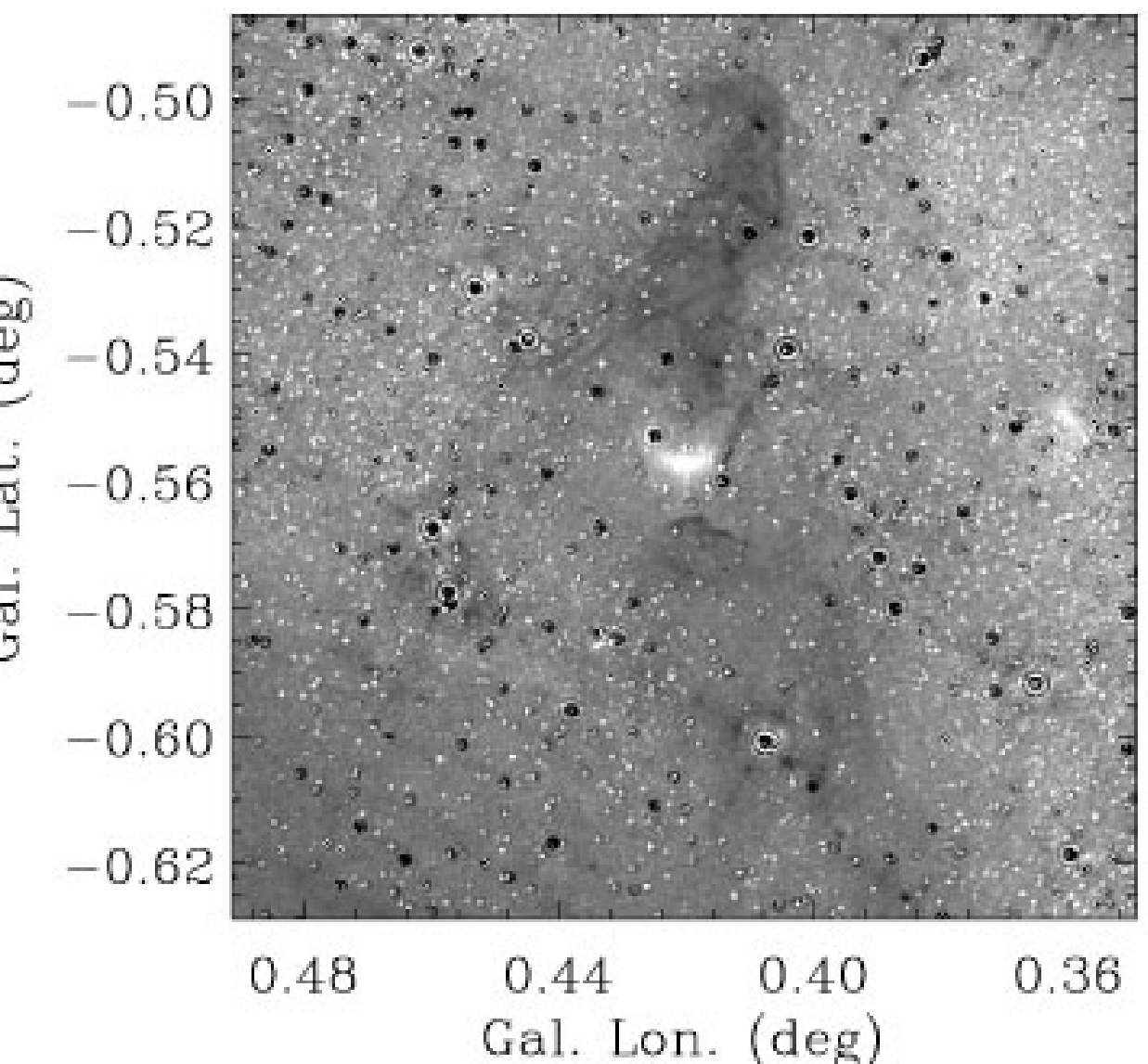}&
\includegraphics[width=2in]{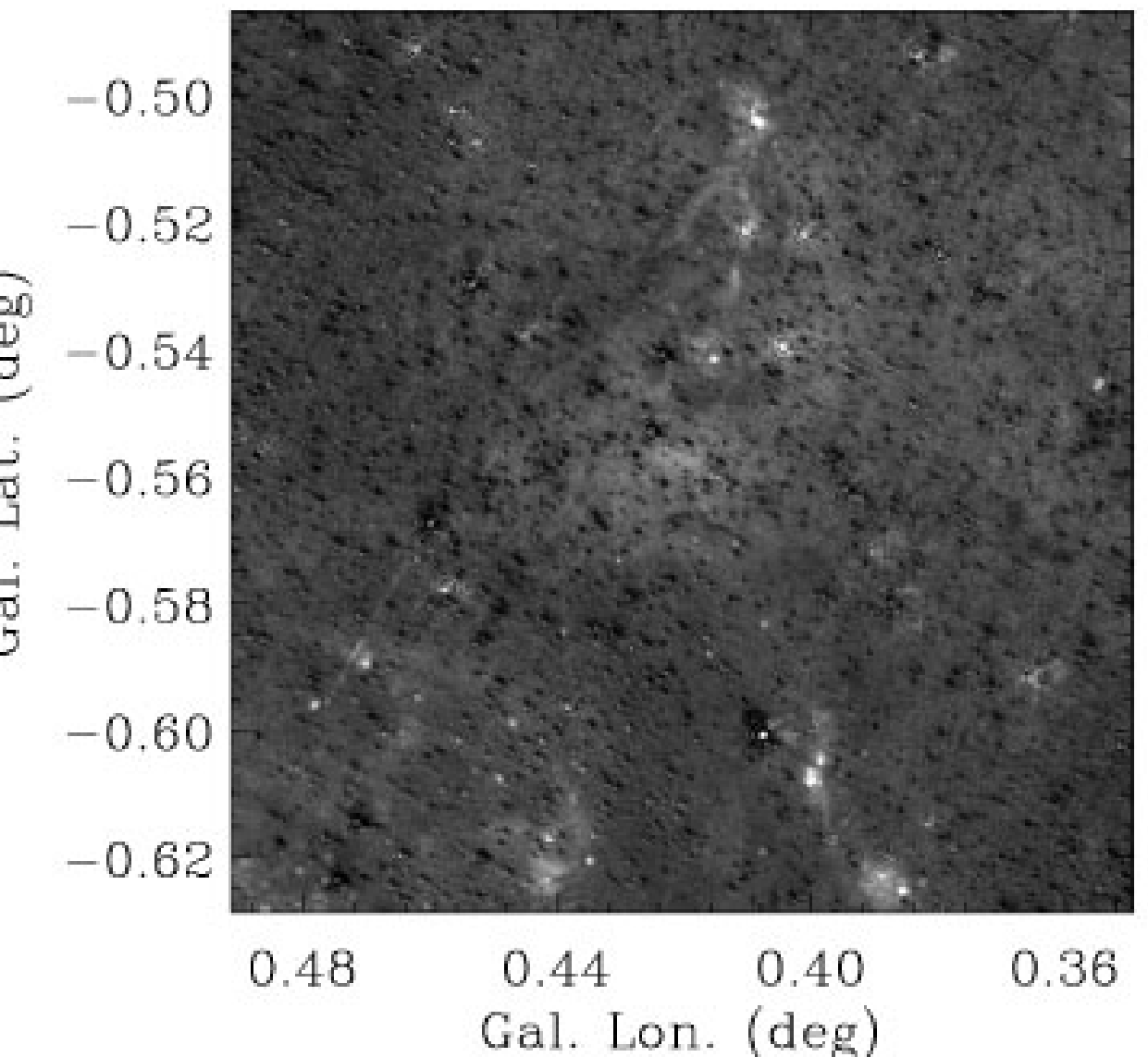}\\
\includegraphics[width=2in]{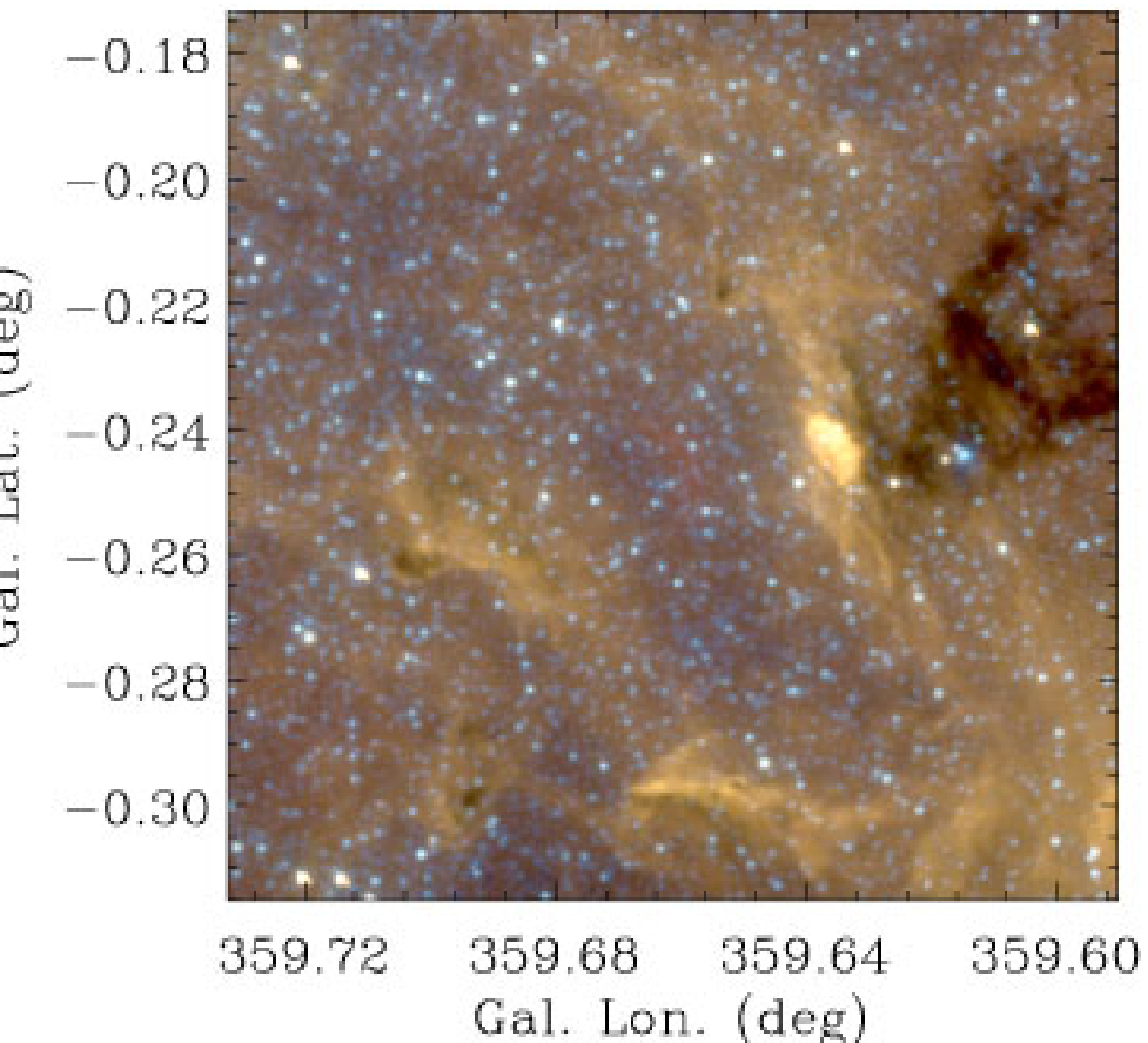}&
\includegraphics[width=2in]{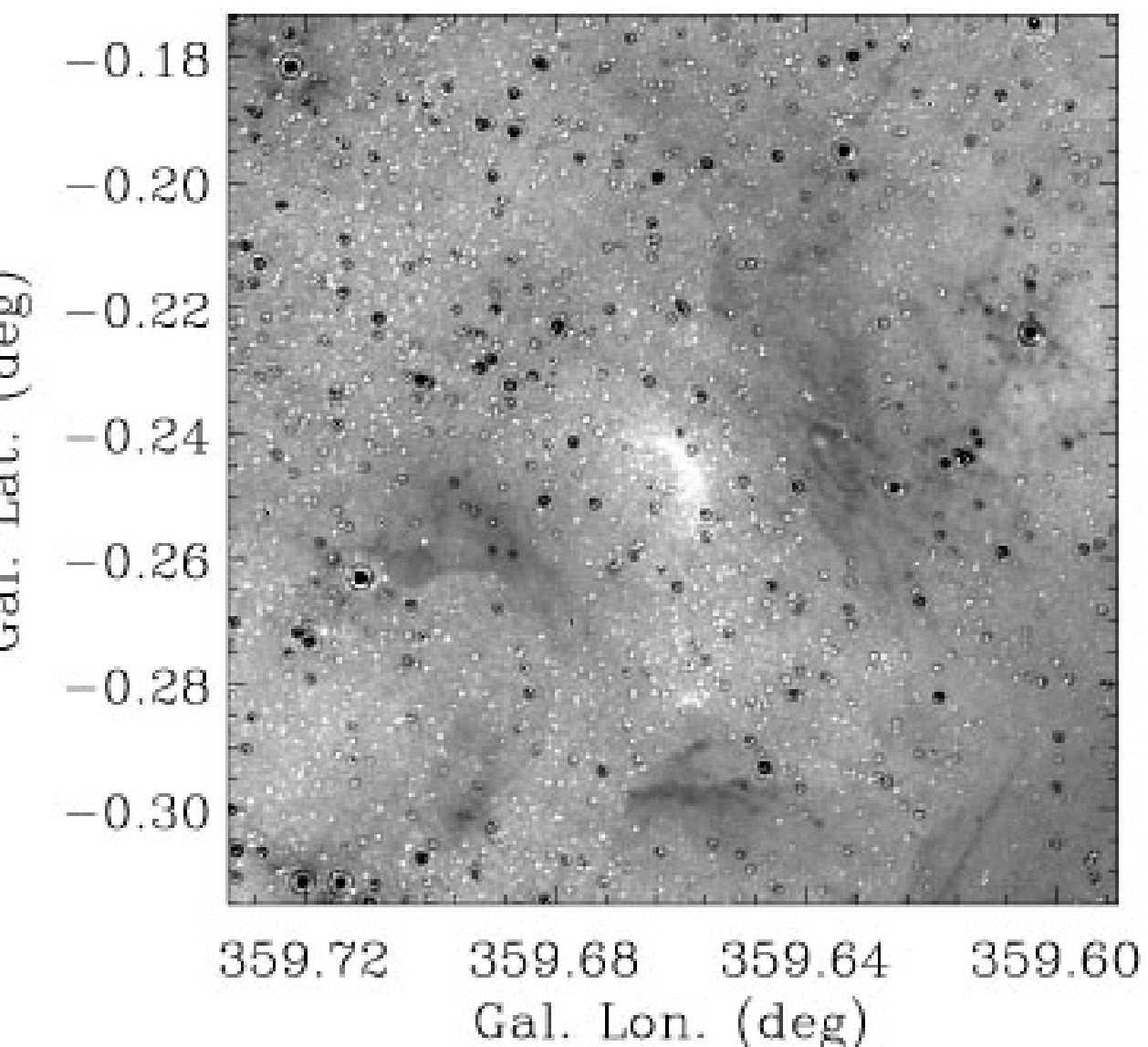}&
\includegraphics[width=2in]{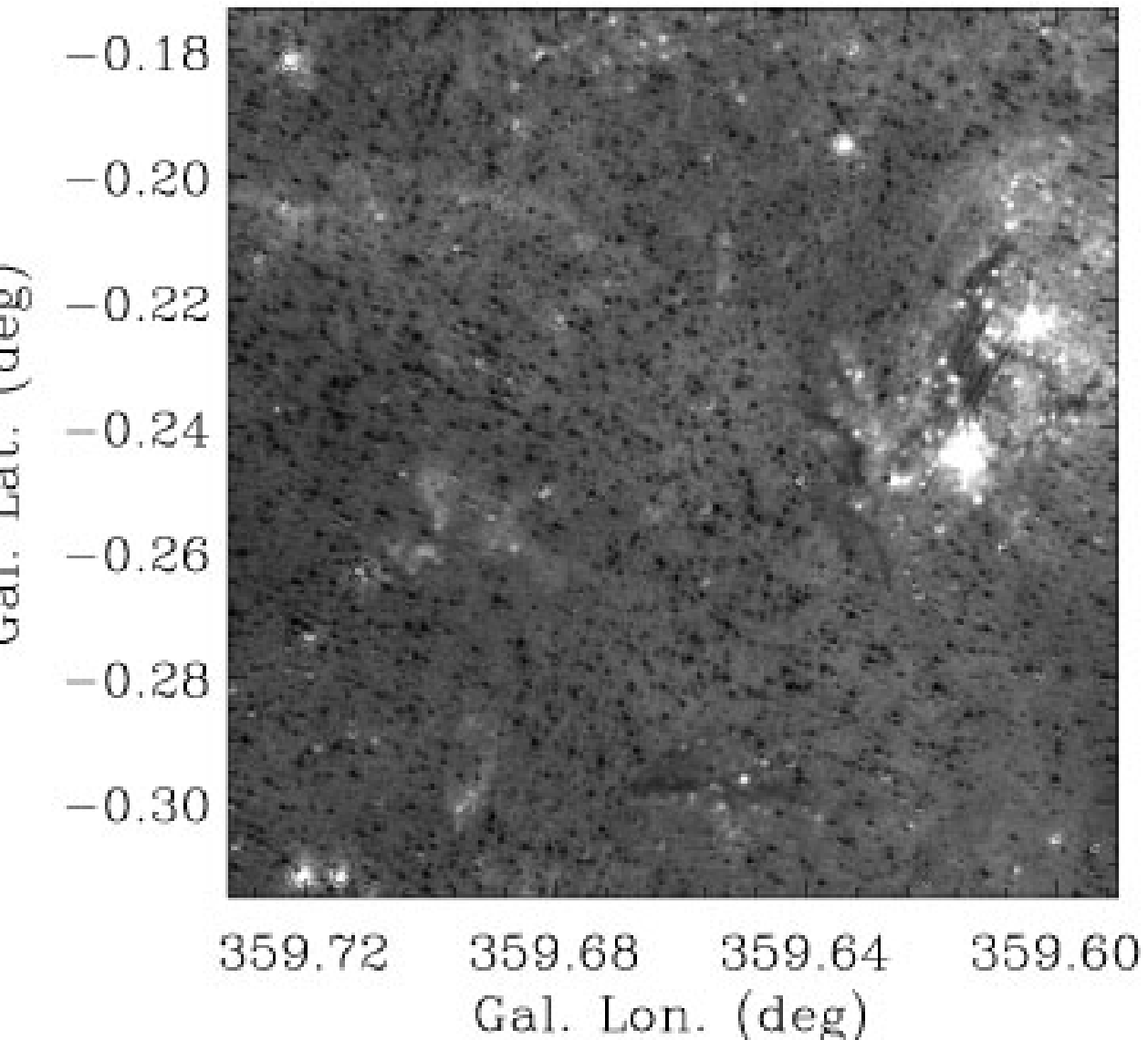}\\
\end{tabular}
\caption{Unusually red regions at high latitudes. IRAS 17456--2850 (top row) 
and IRAS 17425--2920 (bottom row). Left column shows 4.5, 5.8, 8 $\micron$ emission
logarithmically scaled as blue, green, and red respectively. The center column shows 
$I_{ISM}(8\micron)/I_{ISM}(5.8\micron)$ linearly scaled in the range [2.0, 4.0]. 
The right column shows $I(4.5\micron)/I(3.6\micron)$ linearly 
scaled in the range [0.5, 1.5]. 
(In the middle and right columns, lighter shades indicate redder colors.)
Both these IRAS sources are faint and extended, but 
very red in $I_{ISM}(8\micron)/I_{ISM}(5.8\micron)$ colors.
\label{red_regions2}} 
\end{figure}

There are only two extended ($\sim20"$) emission sources $>0.1^{\circ}$ away from 
the Galactic plane that show strong 8 $\micron$ excesses (Figure \ref{red_regions2}). 
Unlike the bright compact H II regions, 
these sources are not very bright at 8 $\micron$, and thus even fainter or absent 
at 5.8 $\micron$. 
However, both brighten rapidly at longer wavelengths
and were detected as IRAS point sources at 25 $\micron$ (\object{IRAS 17456-2850} and 
\object{IRAS 17425-2920}) and are visible in the MSX
band E (21.3 $\micron$) images. Both are located near (in projection) 
extended star forming regions (RCW 137 in the case of IRAS 17425--2920) 
with strong cirrus emission and IRDCs. Both are slightly 
extended and bowed or arced. Neither source contains an obvious point source, nor has a
known energizing star in the crowded field. IRAS 17425--2920 does not appear
in the 1616 GHz radio maps of the Sgr C region (Liszt \& Spiker 1995).
The characteristics and the locations of these 
sources on the outskirts of active star forming regions suggest 
that these sources represent a very early stage of star formation.
Alternately, they could represent local heating by a fully-formed star 
that is interacting with an otherwise quiescent 
molecular or neutral cloud. A third option is that these objects indicate
some sort of shock phenomenon and the unusual colors are the result of strong
line emission. For example, the candidate supernova remnant SSTGFLS J222557+601148 
exhibits only line emission with no continuum (Morris et al. 2006). 
This particular object is bright in the MIPS 24 $\micron$ band by virtue of
a strong [O IV] 26 $\micron$ line, but invisible to IRAC. In the Cas A 
SNR, [Ar II] at 6.99 $\micron$ can contribute 40\% of the emission 
in the IRAC 8 $\micron$ band (Ennis et al. 2006). 
Outside of our survey area at $l = 7\arcdeg$, Rho et al. (2006) detect a similar source in 
the IRAC and MIPS observations of the Trifid Nebula (Rho et al. 2006). 
The region they call ``Trifid Junior'' lies the north of the H II
region and the reflection nebula. It is a small arc of emission at 24 $\micron$ which 
can only barely be seen at 8 $\micron$ and is absent at 5.8 $\micron$. The general 
appearance of the local ($\sim5'$) vicinity is very similar to that of
IRAS 17425--2920. 

\subsection{Molecular Emission: A Dearth of 4.5 $\micron$ Excess Sources}
\begin{figure}
\plottwo{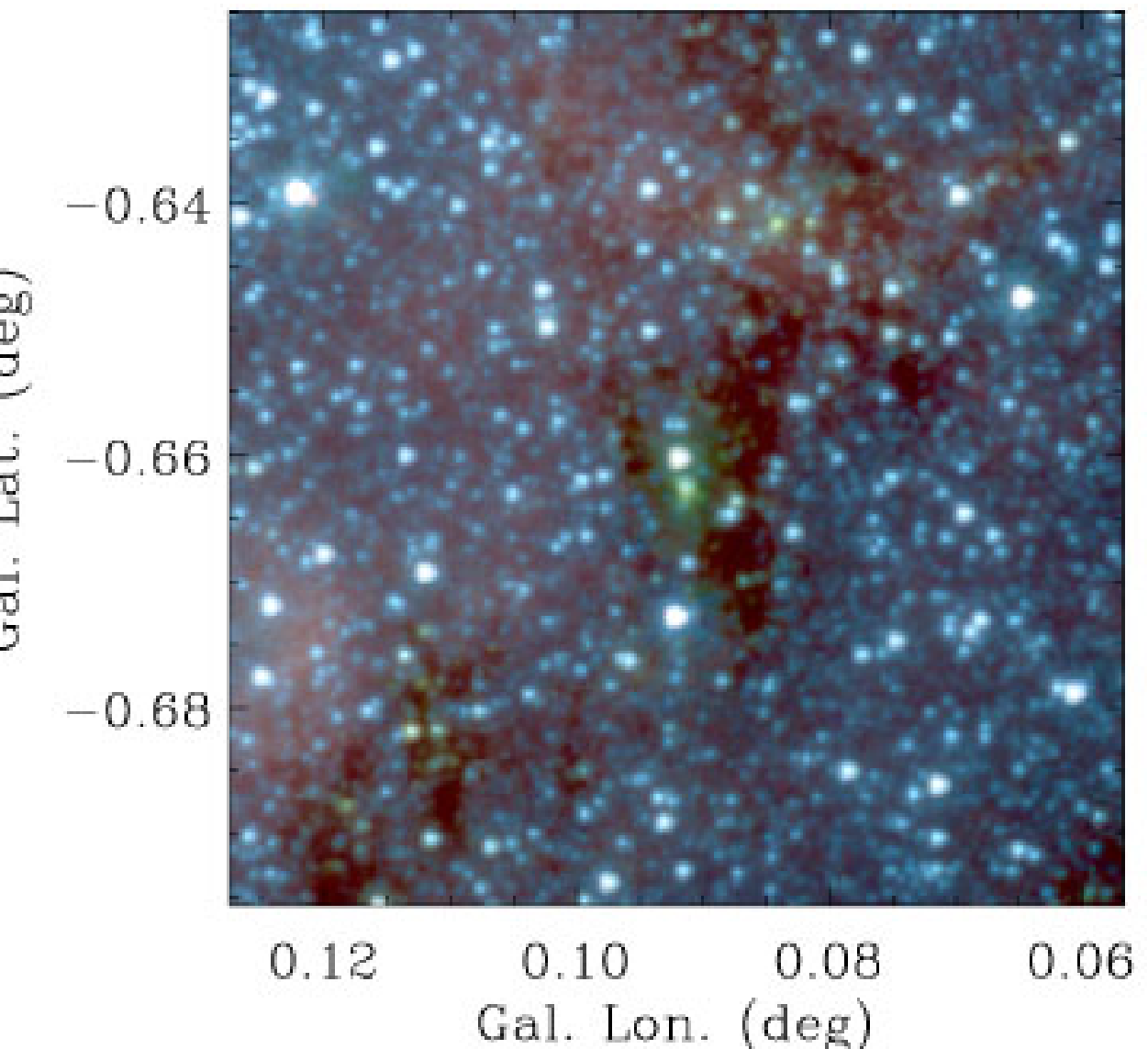}{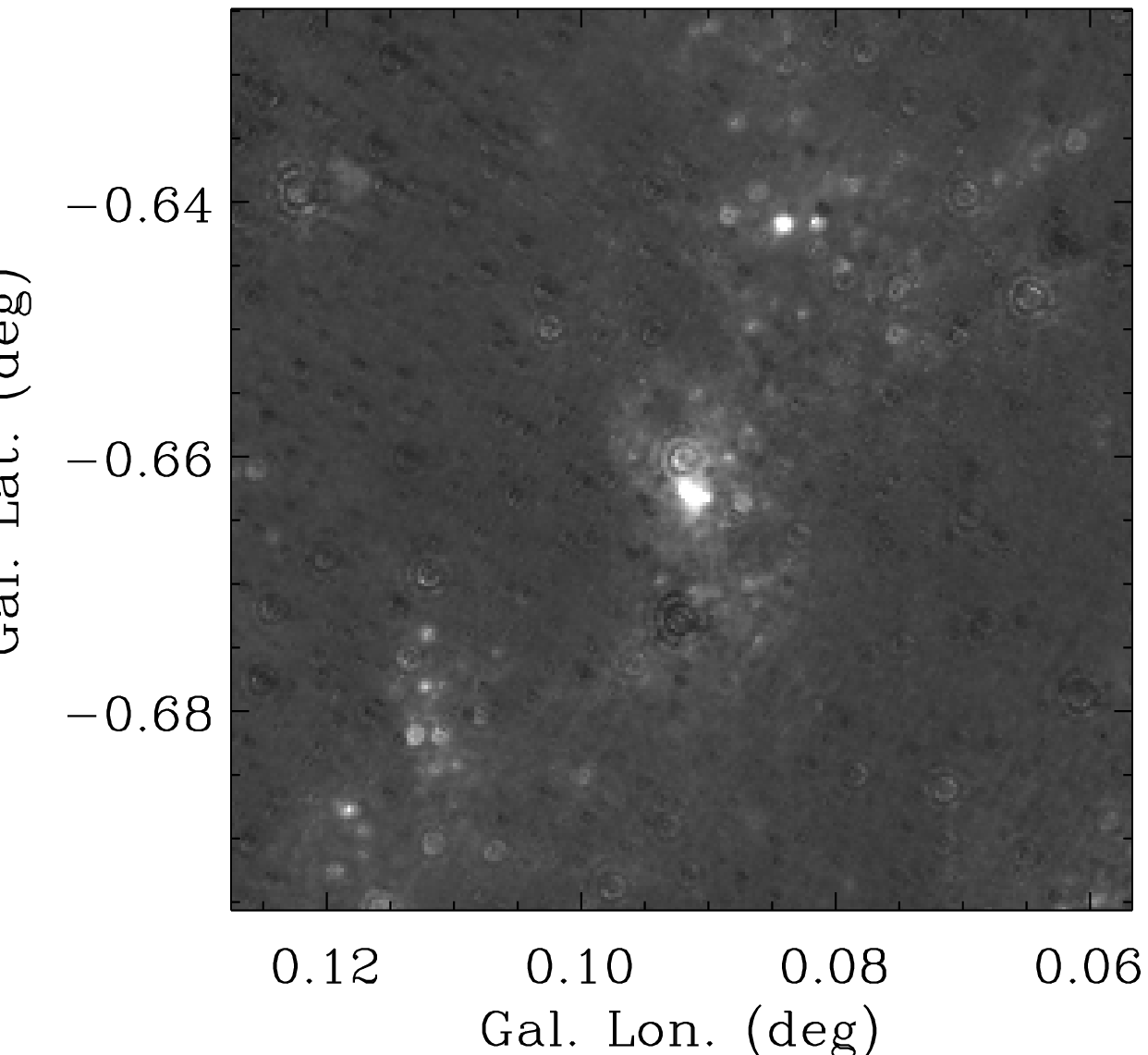}
\caption{The left panel shows 3.6 (blue), 4.5 (green), and 5.8 (red) $\micron$ emission
centered on 2MASS 17482532-2911590. 
There is a slightly extended diffuse feature just $\sim10"$
south of the point source. This extended emission is relatively strong at 4.5 $\micron$ 
as indicated by the greenish color, and a high $I(4.5\micron)/I(3.6\micron)$ ratio
(shown in the right panel on a linear range of [0,3]). This color is often an indication
of molecular emission in outflows or Herbig-Haro objects associated with very young stars.
\label{fig_jet}} 
\end{figure}

Molecular outflows from young stellar objects are often dramatically highlighted
in IRAC images because of CO and H$_2$ emission lines that 
are especially strong in the 4.5 $\micron$ IRAC band (e.g. Smith et al. 2006). 
This wavelength is usually depicted as green in multi--color images, and 
such outflows are therefore recognized by their green colors in such images. 
While some star forming regions can 
contain dozens of outflow sources (e.g. Orion, DR21, NGC 1333),
there is only a single small source that is strong at 4.5 $\micron$ {\it and} 
has a jet--like morphology within our GC survey (Figure \ref{fig_jet}).
There are dozens of other sources that also have 4.5 
$\micron$ excesses, but these are unresolved or only slightly diffuse,
and usually within or near more extended regions of bright 8 $\micron$ emission.
Many of these may be outflows from young stellar objects that 
are too distant or intrinsically small to be resolved. Further discussion
of these sources can be found in Yusef-Zadeh et al. (2007a).

In several of the large, more local star--forming regions at higher
Galactic latitudes there are broad diffuse regions of relatively 
enhanced 4.5 $\micron$ emission [e.g. near $(l, b) = (0\fdg48, -0\fdg67)$]. 
Rather than being an indication of molecular emission, these may be cases
where we spatially resolve an H II region from its surrounding PDR. The 
recent study of M 17 by Povich et al. (2007) shows that the absence of PAH emission
within the H II region where a strong EUV radiation field is present is well 
correlated with an apparent excess at 4.5 $\micron$. Thus the excess is really 
only a decrease in the strength of the PAH features that dominate the 5.8 and 8 
$\micron$ emission (and to a lesser degree 3.6 $\micron$ emission) 
in the surrounding PDR.

\subsection{Unseen Objects}
There are 6 known SNRs 
(Sgr A East, G0.3+0.0, G0.9+0.1, G1.0--0.1, G359.0--0.9 and G359.1--0.5; 
Green 2006) that lie at least partially within
our survey regions, but at 3.6 -- 8 $\micron$ none of these are visible by either their intensity 
or color. This is not unexpected given the high degree of confusion and the fact that SNRs
are often difficult to distinguish in the infrared even with $2''$ spatial resolution.
A thorough examination of the IRAC GLIMPSE survey ($10\arcdeg < |l| < 65\arcdeg$) 
by Reach et al. (2006) found IRAC counterparts 
for only 18 of 95 SNRs. Those that were detected are undergoing strong interactions
with dense, often molecular, clouds. The lack of detections in our survey means that none of 
the 6 SNRs toward the GC is undergoing a strong enough interaction to be visible above the 
brightest background confusion in the Galaxy. The southern portion of IC 443 is a 
well--studied example of a SNR / molecular cloud interaction (e.g. Cesarsky et al. 1999).
Emission with the same surface brightness as IC 443 (AORID = 4422912 and 4423168) 
could be distinguished at higher latitudes, but would be lost in the confusion
at $|b| \lesssim 0\fdg2$. Fainter examples of shocked molecular emission 
such as W44 (Reach et al. 2006) would likely be lost in the confusion at any location
within our survey.

The enigmatic infrared source known as AFGL 5376 does not show any distinct emission at 
any IRAC wavelength, despite being a bright extended source at $\sim24\ \micron$ 
wavelengths (Uchida et al. 1990, 1994; Law 2007). We do note that there are several 
small IRDCs and filaments (in the vicinity of 
$(l,b) = (0\fdg7244,-0\fdg4759)$) that can been seen in extinction 
against the 24 $\micron$ emission of AFGL 5376 as well 
as the general 3.6 -- 8 $\micron$ background.

There is also no particular correlation between the diffuse infrared emission and 
the nonthermal radio filament and threads that are present
in the Galactic center region (Yusef-Zadeh et al. 2004; 
Stolovy et al. 2006, 2008 in preparation).

%%%%%%%%%%%%%%%%%%%%%%%%%%%%%%%%%%%%%%%%%%%%%%%%%%%%
\section{Discussion} 

Overall, the GC exhibits homogenous IRAC colors that are 
quite similar to those of other regions of the Galactic plane.
Such a uniform $I_{ISM}(8 \micron)/I_{ISM}(5.8 \micron)$ color for the 
diffuse emission indicates that there is little reddening 
associated with extinction at these wavelengths, 
despite evidence for highly variable extinction in the form of IRDCs
and extinction measurements made at shorter wavelengths (e.g. 
Gosling et al. 2006). 
This is expected given the flattening of Galactic extinction curve 
(Lutz et al. 1996; Indebetouw et al. 2005; Flaherty et al. 2007).
Even if the extinction did produce reddening, it would be extremely unlikely 
for the line of sight reddening to be correlated with variation in 
the intrinsic color of the emission to produce such an overall uniform color.

\begin{figure}
\plotone{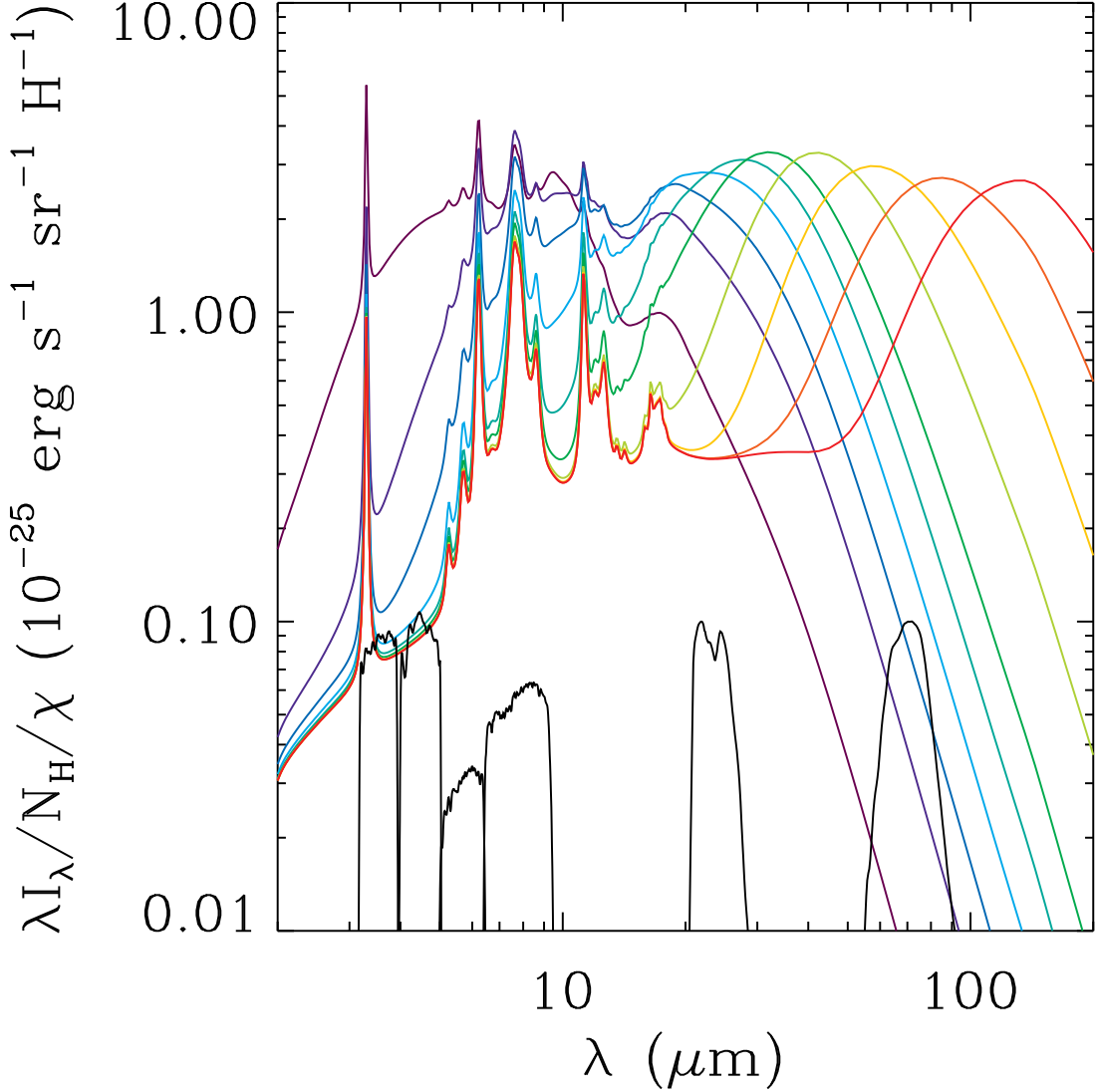}
\caption{The colored curves show the emission spectra for a selected dust grain
model (``BARE-GR-FG'' from Zubko et al. 2004) subjected
to radiation fields of 
$\chi = 10^{[0,1,2,3,4,4.5,5,5.5,6,7]}$ times the local ISRF.
The black lines indicate the filter responses (arbitrarily normalized) 
for the 4 IRAC channels, and for the MIPS 24 and 70 $\micron$ channels.
The PAH spectrum is unchanged with radiation field strength, 
and hence the IRAC flux ratios remain unaltered until the larger, 
roughly equilibrium temperature grains get so warm that they encroach 
on the 8 $\micron$ IRAC band.
Note that the continuum, not PAH emission bands, 
dominates the 3.6 and 4.5 $\micron$ IRAC filters. The 3.3 $\micron$ PAH feature
contributes $<30\%$ of the total flux in the IRAC 3.6 $\micron$ channel for these models.
\label{fig_ZDAmodels}} 
\end{figure}

The uniform colors are also evidence that the emission spectrum is largely independent 
of the local and global locations and environments. This strongly suggests that
emission from polycyclic aromatic hydrocarbons (PAHs) dominates the diffuse
ISM emission (5.8-8 $\micron$), all the way to the Galactic Center. The relative strengths 
of PAH emission bands that fall within the IRAC 5.8 and 8 $\micron$ 
filter bandpasses are unaffected by the strength of the radiation field heating
the grains over a wide range of intensities (Li \& Draine 2001). Even as 
the interstellar radiation field strength increases toward the GC, there is no 
evident change in the color of the PAH emission. Figure \ref{fig_ZDAmodels} shows model 
spectra of the ISM emissivity normalized to the strength of 
the interstellar radiation field (Mathis et al. 1983), 
as characterized by a scale factor $\chi$. The dust model used 
here is composed of PAHs and bare graphite and silicate grains
(the ``BARE-GR-FG'' model of Zubko et al. 2004). 
This model is similar to those of Li \& Draine (2001), but is derived 
via additional constraints provided by elemental abundances. 

For the standard ISRF ($\chi = 1$), most emission at $\lambda < 60\ \micron$
is from very small grains ($\lesssim 0.01\ \micron$) and PAHs. 
These grains have a low heat capacity 
and are heated to high temperatures by absorption of a single photon. 
They radiate that energy quickly and cool below equilibrium temperatures
before the next energetic photon hits. Thus, for these stochastically heated
particles, the grains do not get any warmer as the radiation field increases; 
they only go through the same temperature spikes more frequently.
Therefore, the spectrum normalized by $\chi$ remains roughly constant.
This is especially evident at $\lambda < 10\ \micron$.
Larger grains ($\gtrsim 0.01\ \micron$) 
have higher heat capacities and absorb photons frequently 
enough that they radiate at their equilibrium temperatures.
For the local ISRF ($\chi = 1$), this thermal emission peaks at
$\sim 150$ $\micron$. As $\chi$ increases, the larger grains get warmer and 
the thermal emission moves to shorter wavelengths. Eventually, when 
$\chi \gtrsim 10^4$, the thermal emission begins to enter the 8 $\micron$
IRAC band. At this point the IRAC colors can be affected.

\begin{figure}
\plottwo{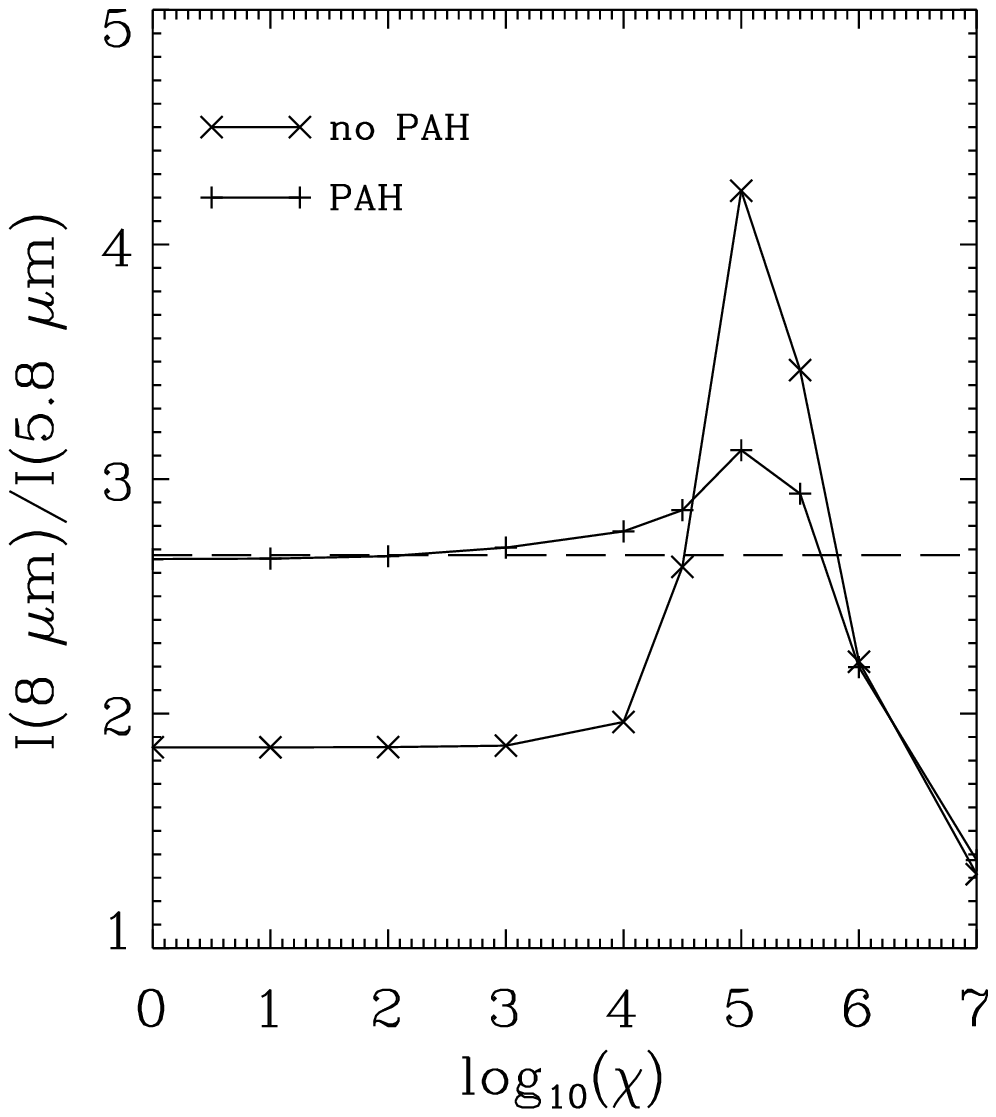}{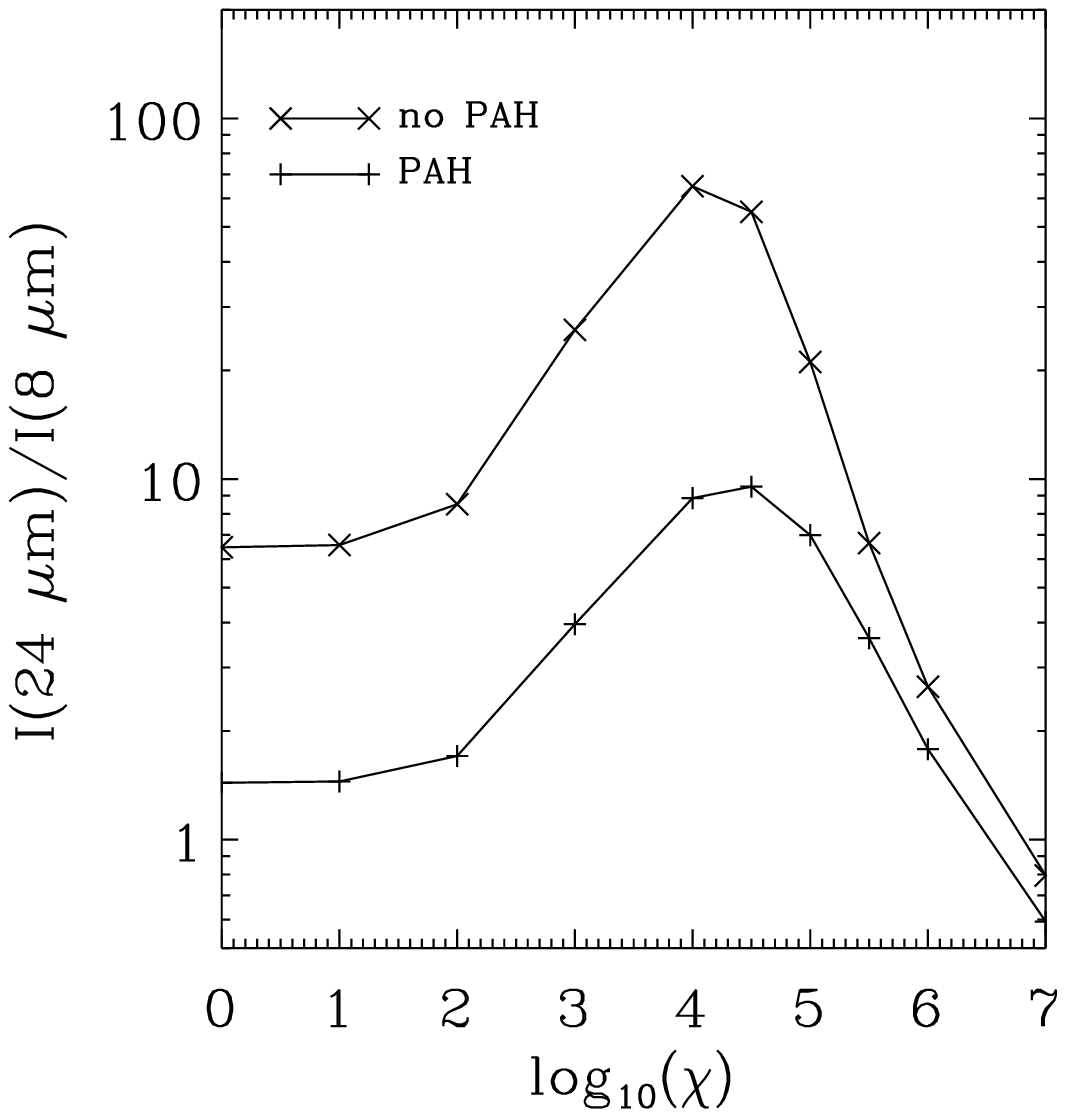}
\caption{(left) This figure quantifies the $I(8\ \micron) /I(5.8\ \micron)$ ratio
for the dust models shown in Figure \ref{fig_ZDAmodels}. 
The dashed line is the median ratio over the
whole GC region.
(right) This plot shows the corresponding the $I(24\ \micron) /I(8\ \micron)$ ratio. 
Involving longer wavelengths, this color is more sensitive to changes 
in the cooler, larger dust grains and lower values of the radiation
field $\chi$.
\label{fig_ZDA_ratios}} 
\end{figure}

In Figure \ref{fig_ZDA_ratios} we plot $I_{ISM}(8 \micron)/I_{ISM}(5.8 \micron)$ predicted
by the models of a function of radiation field strength. 
The figure shows the ``BARE-GR-FG'' model is consistent with the observed
$I_{ISM}(8 \micron)/I_{ISM}(5.8 \micron)$ mean color as long as $\chi \lesssim 10^4$.
It is also evident that while the color initially reddens as
the thermal emission from the hot dust enters the 8 $\micron$ band,
the color begins getting bluer when the thermal emission pushes into the 
5.8 $\micron$ band as well. However, the reddest colors attained by this 
model (at $\chi \sim 10^5$) are not nearly red enough to explain the 
unusually red regions that are we identify in the GC region (\S 3.2).
Alternatively, if the PAH component is removed from the ``BARE-GR-FG'' model,
then at $\chi \sim 10^5$ the color of the emission can be sufficiently 
red to match the observations. 
At $\chi \lesssim 10^4$, the colors of the emission would be much bluer
than observed if PAHs were generally absent from the ISM in the GC.
Qualitatively similar results are obtained 
from the Draine \& Li (2007) dust models, which include a 
parameter ($q_{PAH}$) that characterizes the relative abundance of PAHs.

The red shell around the Arches Cluster provides a nice example of
a region with a locally strong radiation field. Figer et al. (2002)
have estimated that luminosity of the Arches Cluster as $10^{7.8} L_{\sun}$.
The extended red shell has a radius of $\sim 100"$ or 3.9 pc.
Thus the energy density at the outer edge of this shell is 
$\sim 4.4 \times 10^{-9}$ erg cm$^{-3}$. Compared to the nominal 
strength of the local ISRF, $7 \times 10^{-13}$ erg cm$^{-3}$ 
(Spitzer 1978), we find that $\chi \sim 6300$ at the edge of the shell.
This value of $\chi$ indicates that we should indeed expect to see 
emission with red $I_{ISM}(8 \micron)/I_{ISM}(5.8 \micron)$ colors in the vicinity 
of the Arches Cluster, and that the region of red emission cannot 
be much larger than what is observed. 

The red emission region near the Quintuplet Cluster is related to a 
similarly enhanced radiation field. This region corresponds to the 
portion of the ``Sickle'' that is nearest to the Quintuplet Cluster.
Spectroscopy of this region by Simpson et al. (1997, 2007) reveal 
gas in a state of high excitation, consistent with radiation fields 
of $\chi > 10^3$. Calculations by Rodr\'iguez-Fern\'andez et al.
(2001) indicate that the Quintuplet Cluster will fill a smaller volume 
with a strong radiation field than the Arches Cluster. 
This is in agreement with the smaller extent of the red emission 
in the vicinity of the Quintuplet Cluster
compared to the Arches Cluster (Fig. \ref{red_regions}). 
In particular, the extent of the radiation field strong enough to cause 
redder $I_{ISM}(8 \micron)/I_{ISM}(5.8 \micron)$ colors does not reach to the 
(Galactic) southern end of the ``Radio Arc Bubble''. This is 
consistent with the fact that while $20-25 \micron$ observations 
(IRAS, MSX, MIPS) show a complete bubble, the portions more distant 
from the Quintuplet cluster do not exhibit unusually red 
$I_{ISM}(8 \micron)/I_{ISM}(5.8 \micron)$ colors. Comparison of the 
spectroscopic results of Simpson et al. (2007) indicates that the 
$I_{ISM}(8 \micron)/I_{ISM}(5.8 \micron)$ color is fairly well correlated 
(correlation coefficient $r > 0.6$) with 
the continuum intensity at $\lambda > 10\ \micron$, and with 
H 7-6 12.37 $\micron$, [Ne II] 12.81 $\micron$, [P III] 17.89 $\micron$, 
[S III] 18.71, 33.48 $\micron$, and [S IV] 10.51 $\micron$. 
The correlation of the $I_{ISM}(8 \micron)/I_{ISM}(5.8 \micron)$ color with the 
$\lambda > 10\ \micron$ continuum color, or with line ratios 
is weaker. An exception is that correlation with 
[Si II] 34.83 $\micron$ / [Fe II] 26 $\micron$ 
(or 17.94 $\micron$) is quite strong, $r > 0.75$.
This is rather surprising as both species have similar ionization potentials
and would be similarly affected by depletion into (or destruction of) dust grains.

%%%%%%%%%%%%%%%%%%%%%%%%%%%%%%%%%%%%%%%%%%%%%%%%%%%%
\section{Summary}

Our {\it Spitzer}/IRAC survey of the Galactic Center 
(Stolovy et al. 2006, 2008 in preparation) 
reveals prominent 5.8 and 8 $\micron$ emission from the ISM in the 
central $\sim$ 200 pc of the Galaxy in unprecedented detail. 
The infrared emission is loosely associated with molecular
emission from the CMZ, and several times brighter than 
infrared emission elsewhere in the Galactic plane. In general, the 
$I_{ISM}(8 \micron)/I_{ISM}(5.8 \micron)$ color of the emission is very uniform 
across the CMZ, and similar to that seen in other parts of the Galactic 
plane (e.g. Flagey et al. 2006). 
Clearly we cannot infer too much from a single color, given the 
large number of parameters that can be adjusted for the dust models
used to interpret the emission. However, we can generally conclude 
the predominant uniformity of the observed $I_{ISM}(8 \micron)/I_{ISM}(5.8 \micron)$ 
color in the GC implies 
(1) the emission is dominated by PAHs, 
(2) the general ISRF in the central 200 pc of the Galaxy does not reach extreme 
values ($\chi \lesssim 10^4$), and 
(3) the interstellar extinction law at these wavelengths is 
flat (e.g. Indebetouw et al. 2005; Flaherty et al. 2007) leading to no significant
reddening despite obvious cases of locally high extinction.

There are several extended regions that exhibit much redder than average
$I_{ISM}(8 \micron)/I_{ISM}(5.8 \micron)$ colors despite not being especially bright.
Red regions around the Arches Cluster, the Quintuplet Cluster, and Sgr B1 
are apparently produced by a combination of locally strong radiation 
fields ($\chi \gtrsim 10^4$), and a lack of PAHs (perhaps a consequence
of the strong radiation field). Other very red regions that are more compact 
and further from the Galactic plane appear to be associated with 
the onset of star formation, although they do not appear to contain any
detectable point sources.

\acknowledgments
This work is based on observations made with the Spitzer Space 
Telescope, which is operated by the Jet Propulsion Laboratory, California 
Institute of Technology under a contract with NASA. Support for this 
work was provided by NASA through an award issued by JPL/Caltech.
This research has made use of the NASA/ IPAC Infrared Science Archive, which 
is operated by the Jet Propulsion Laboratory, California Institute of Technology, 
under contract with the National Aeronautics and Space Administration. 
We thank the referee for pointing out the non-negligible influence
of the 4.5 $\micron$ ISM emission in our analysis.

{\it Facilities:} \facility{Spitzer (IRAC)}

\appendix
%%%%%%%%%%%%%%%%%%%%%%%%%%%%%%%%%%%%%%%%%%%%%%%%%%%%
\section{Data Reduction Details}
\subsection{Post--BCD Processing}

The BCD are single frames (exposures) calibrated with the standard calibration data 
applied by the Spitzer Science Center (SSC) pipeline. While on average this calibration is
quite good, in some cases the state of the IRAC detectors may be different from what the 
standard calibration assumes. Thus the BCD can still contain some types of instrumental 
artifacts, and the individual frames may not blend smoothly into a large mosaicked image,
especially when that image incorporates multiple AORs.

One significant artifact in the 3.6 and 4.5 $\micron$ channels is ``column pulldown''.
This is an electronic artifact that induces low intensities in detector 
columns coincident with bright (usually saturated) sources. 
A standard correction was applied to individual BCD frames for the 3.6 and 4.5 $\micron$ channels. 
This greatly mitigates the effect, but does not reduce it to undetectable levels. 
A similar correction for banding was applied to 8 $\micron$ BCD frames 
to eliminate excess intensity in detector rows containing very bright sources. The 5.8
$\micron$ data could have benefitted from such corrections in both detector 
rows and columns. However the character of the artifact is sufficiently different 
from the behavior at 8 $\micron$,
that attempts at correcting 5.8 $\micron$ banding were not successful enough to 
apply them in an automated way. Thus these artifacts remain, and are a source of 
systematic error in particular regions where saturated sources are clustered. 

\begin{figure}
\plottwo{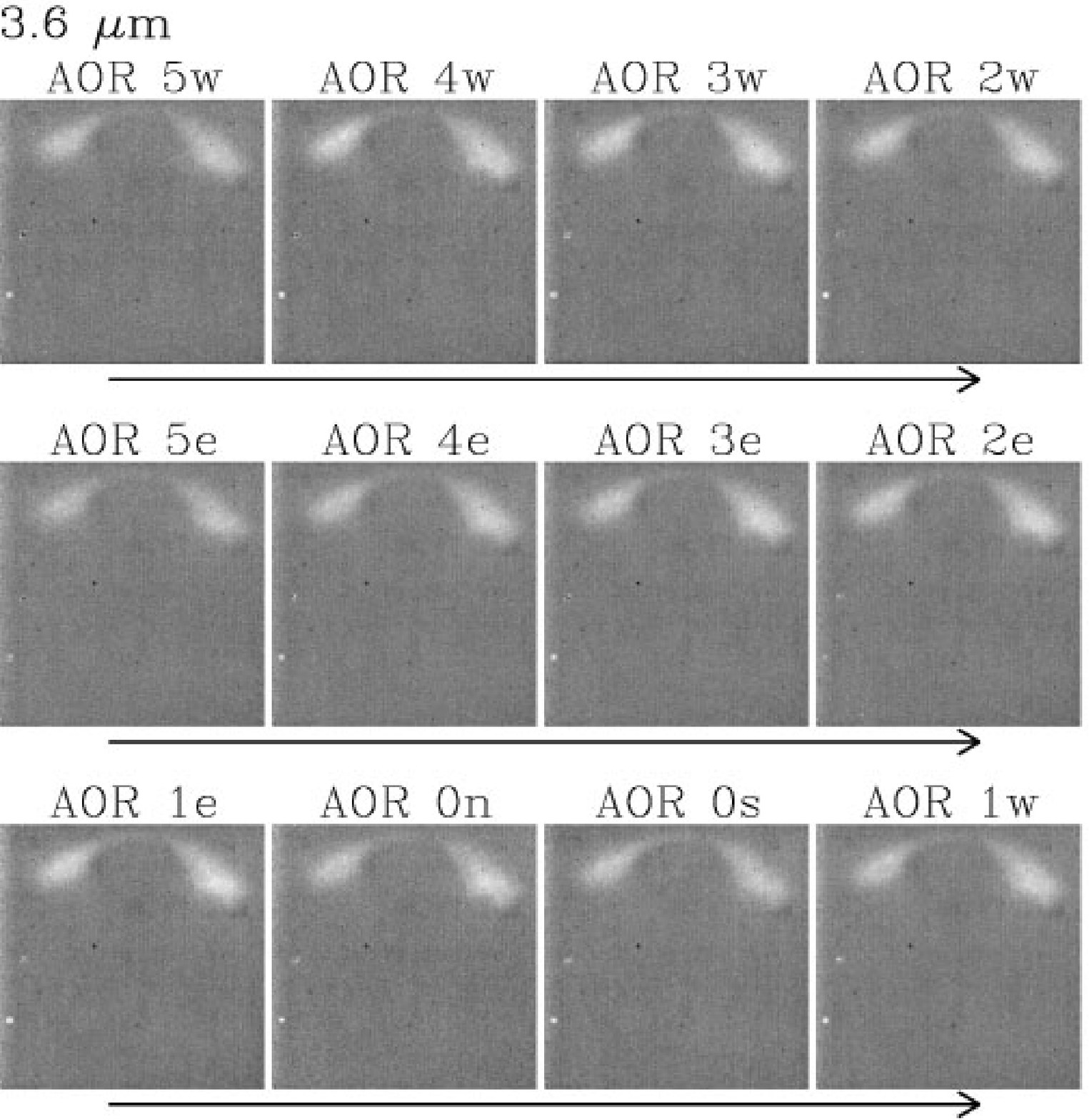}{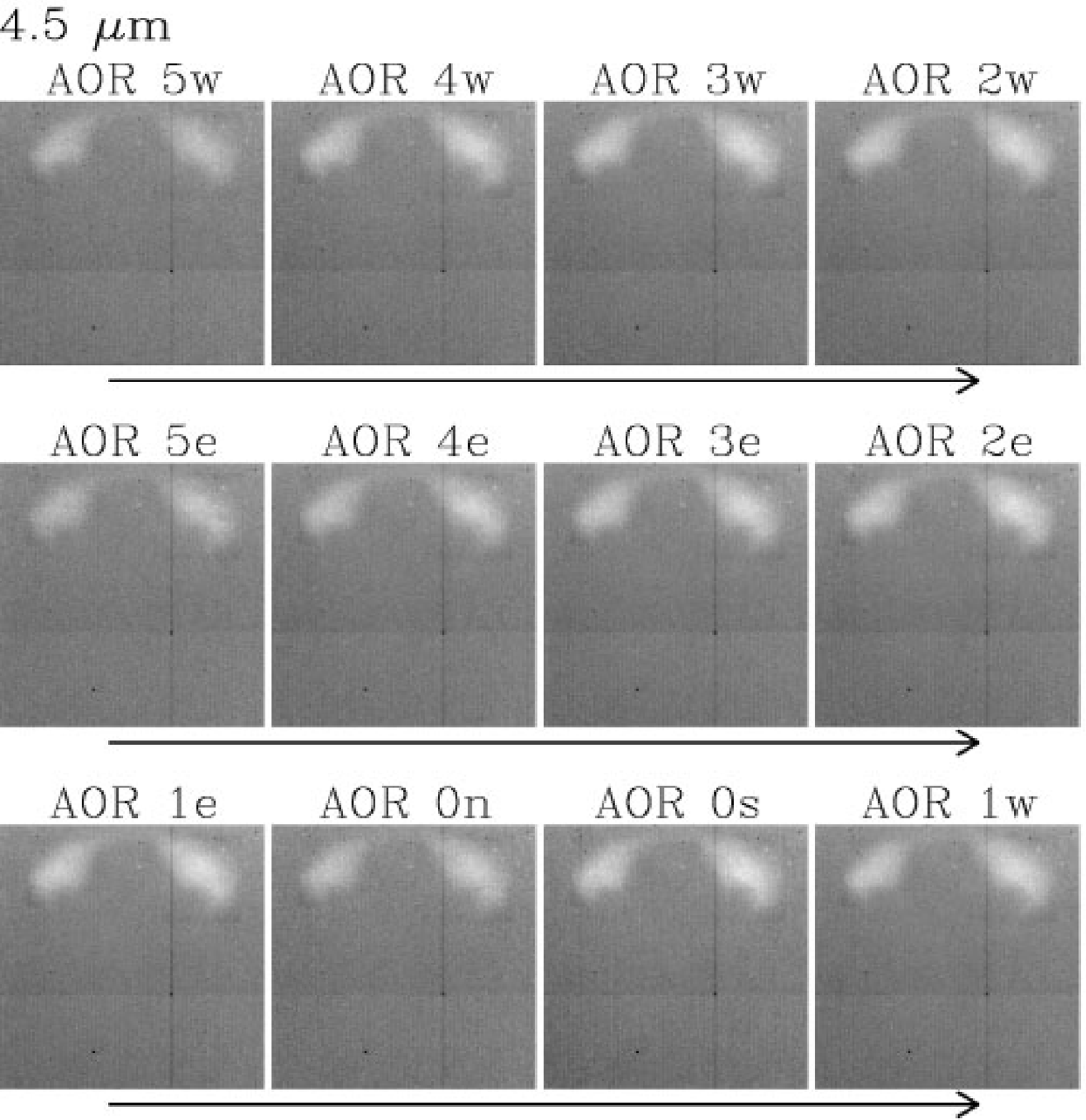}
~~\\
\plottwo{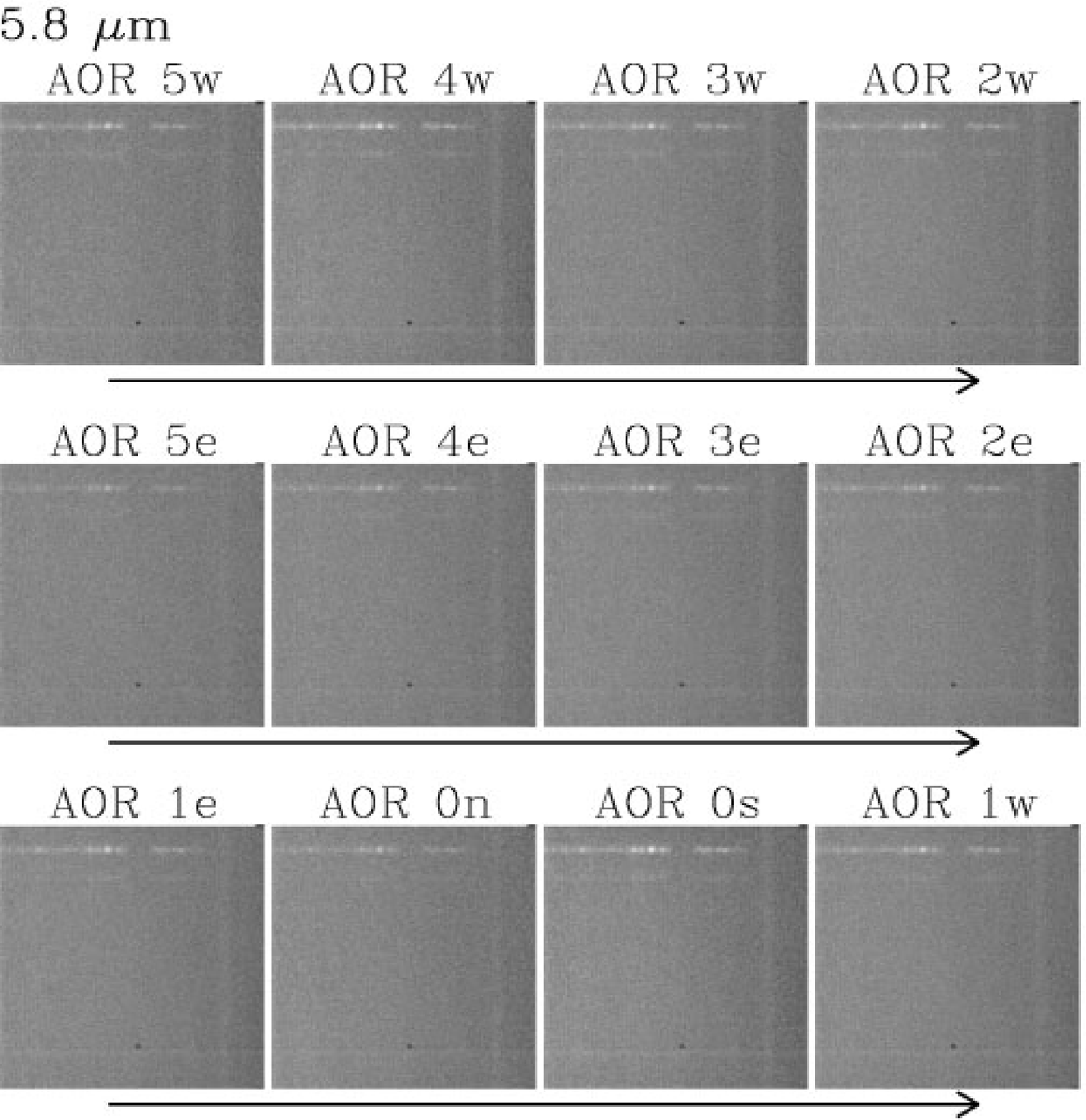}{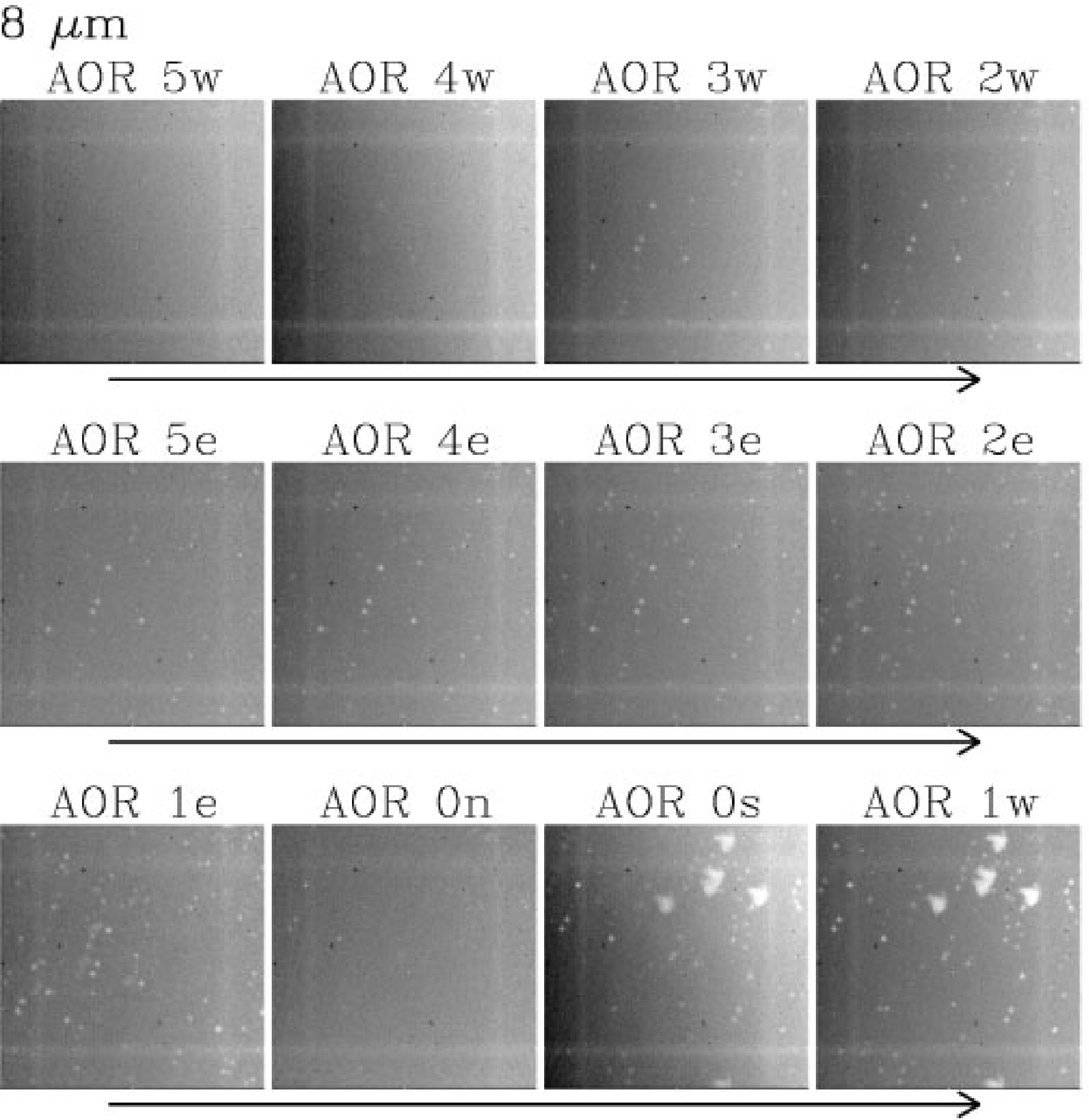}
\caption{(a-d) Images of the offsets ($F^p$) derived for each AOR in a time ordered
sequence. These show the diffuse stray light artifacts that remained in 
the BCD, and the accumulation of long--term latent images at 8 $\micron$. 
The last three AORs (0n, 0s, 1w) were separated from the first by a downlink and 
anneal, which cleared previous long term latent images. The Sgr A region creates 
strong latent images during AOR 0s, which then remain throughout AOR 1w.
Display ranges are [-2,2], [-2,2], [-5,5], [-4,4] MJy sr$^{-1}$ for 3.6 
through 8 $\micron$ respectively.
\label{fig_fp}
}
\end{figure}

\begin{figure}
\plotone{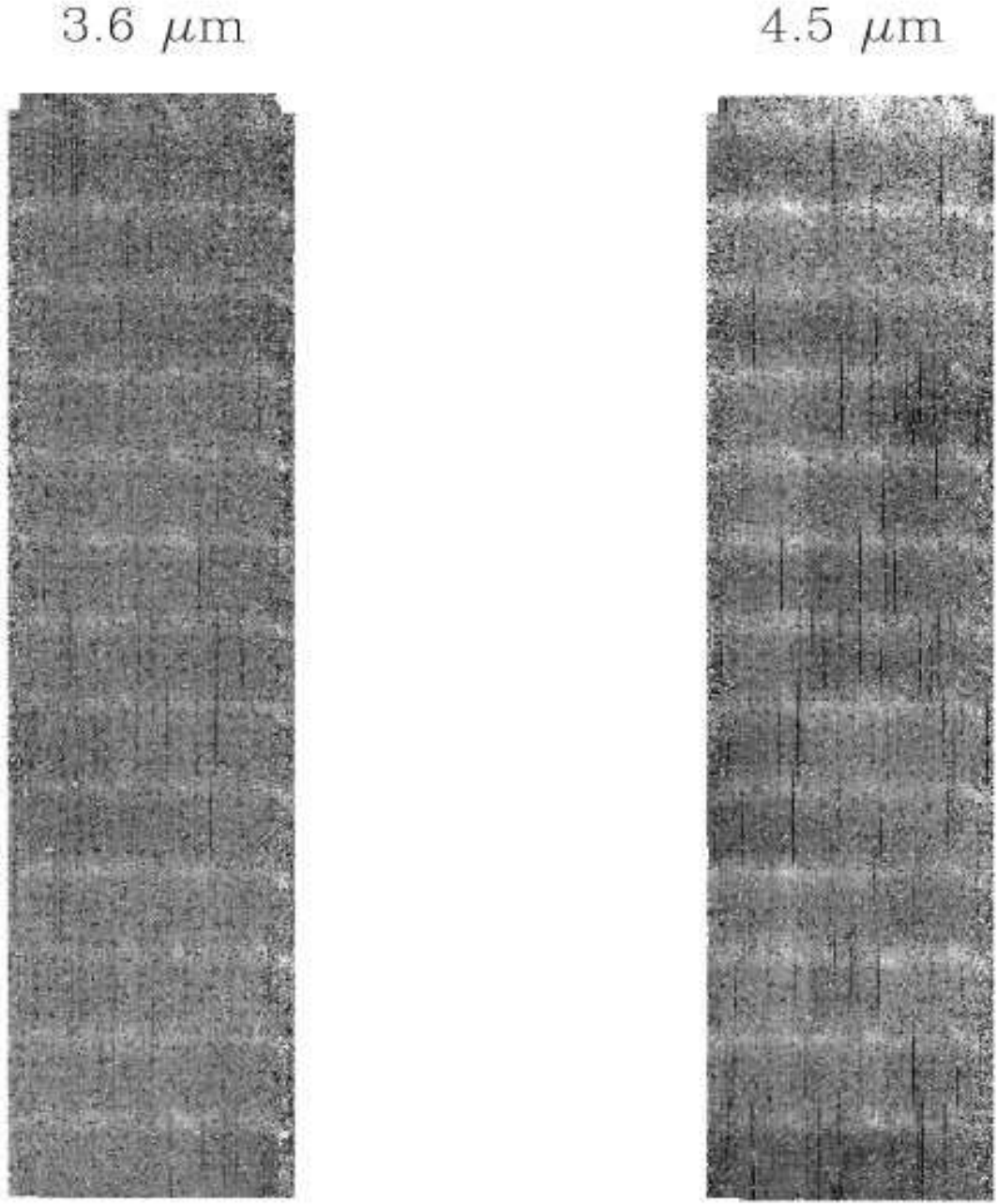}
\caption{Ratios of standard post-BCD mosaics to our self--calibrated BCD mosaics 
for one AOR strip of the survey (AOR key = 13371648) 
at 3.6 $\micron$ (left) and 4.5 $\micron$ (right). Display range = [0.9,1.2].
The light horizontal bands indicate the regions that are affected by the 
diffuse stray light in the post--BCD mosaics. 
These artifacts are present in the ratio of 3.6/4.5 $\micron$ post-BCD
mosaics, but not in ratios of our self--calibrated mosaics.
The dark vertical lines illustrate the changes due to the application of 
the column pull--down correction.\label{fig_rat12}}
\end{figure}

The BCD frames for each AOR were self-calibrated using the procedure 
described by Fixsen et al. (2000). This procedure uses the bright infrared 
background emission (combined zodiacal light and Galactic emission) as a calibration source.
The background is not uniform, but it is stable. Thus, when 
the dithering and mapping place different detector pixels on the same sky location, the 
resulting values for {\it properly} calibrated data should be the same, apart from a random 
noise variation from gaussian detector noise and Poisson photon shot noise. Not all
detector pixels observe a single sky location, but with a large data set such as
ours, if two different detector pixels did not observe a common sky location then there
is an intermediate pixel (or a chain of several such pixels) that has observed at least
one sky location in common with each of the original two pixels. Through such inter-comparisons,
the relative calibrations of each detector pixel can be determined. This is solved
using a linear least--squares algorithm. In the framework of this procedure 
the data ($D^i$, i.e. each pixel of each BCD frame) are modeled 
as the sum of the true sky intensity ($S^{\alpha}$ at sky coordinates $\alpha$),
a fixed detector offset ($F^p$ as a function of detector pixel $p$), 
and a variable offset ($F^q$ a single value for all pixels of frame $q$):
\begin{equation}
D^i = S^{\alpha} + F^p + F^q .
\end{equation}
The least--squares self--calibration allows us to determine the offset terms in the
model (i.e. the proper detector calibration), and then invert Eq. A1 to derived the 
true sky intensity, $S^{\alpha}$ from the data, $D^i$. 
The images of the fixed detector offsets ($F^p$) are shown in Figure \ref{fig_fp} for each 
IRAC channel and each AOR. For the 3.6 and 4.5 $\micron$ data, the obvious feature is the 
diffuse stray light ``butterfly wing'' pattern at the top of the detector. This feature 
represents the integrated stray light of all the sources of the bright unresolved 
background, which is not effectively removed through the standard BCD calibration.
The amplitude of this artifact is small and is not evident directly in the single BCD frames or the mosaics, 
but if left uncorrected it would induce artifacts in the ratio of 3.6 / 4.5 $\micron$ emission as
shown in Figure \ref{fig_rat12}. This figure also illustrates that the column pulldown correction is 
not entirely effective at both 3.6 and 4.5 $\micron$ for the very brightest sources, but the 
pulldown correction usually does improve the data for weaker, more frequent cases. 
The fixed detector offsets for the 5.8 and 8 $\micron$ channels also show (weakly)
the diffuse stray light patterns for these channels (Fig. \ref{fig_fp}). 
Additionally the 8 $\micron$ results clearly show the accumulation of latent 
images from bright sources, 
the clearing of these after an anneal, and the further accumulation of new latent images.
The strong latent images acquired at 8 $\micron$ during AOR 0s, 
are from the central cluster at Sgr A.
The relative strength and extent of the latent image indicates that the Sgr A cluster
is clearly the highest density of bright sources in our survey.

\begin{figure}
\plotone{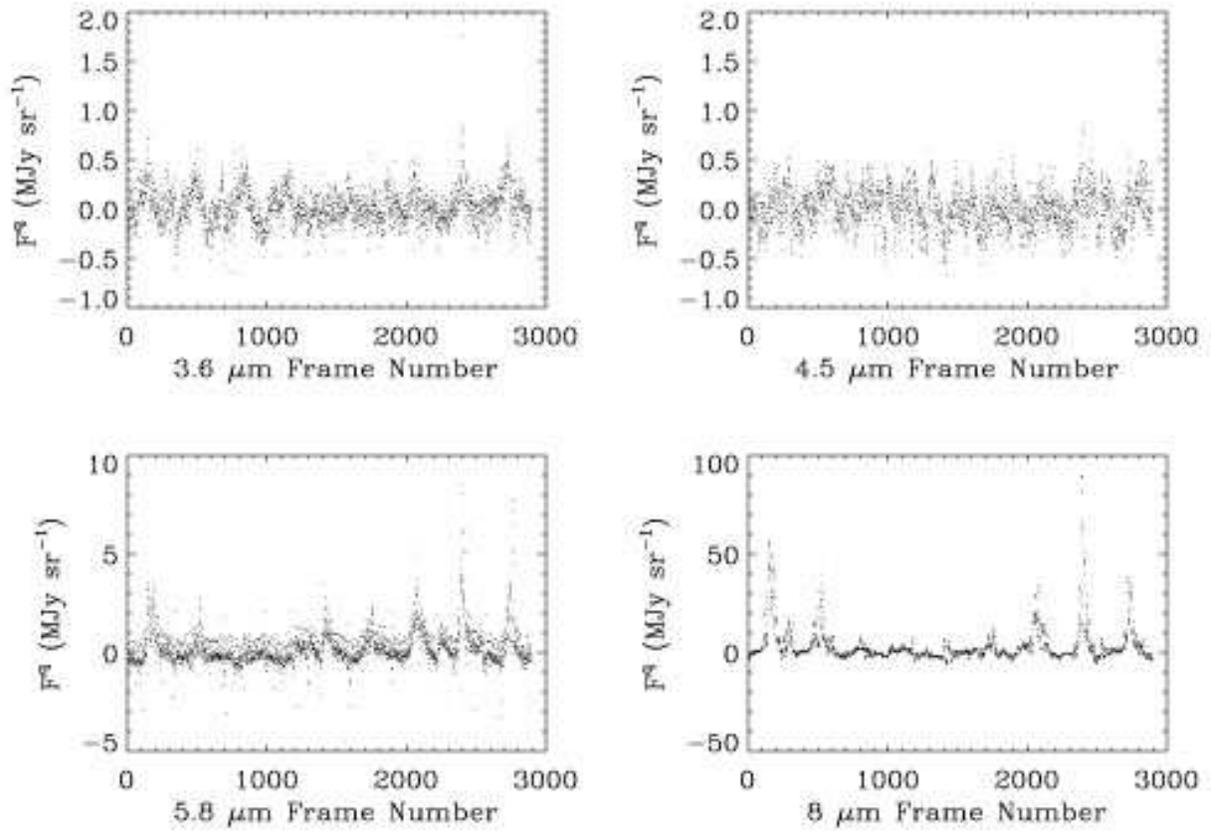}
\caption{Variable offsets per frame ($F^q$) for each channel are plotted 
in a time ordered sequence. The 5.8 and 8 $\micron$ channel offsets are relatively 
large and well--correlated with the local sky intensity, as revealed by
mapping the offsets (Figure \ref{fig_fqmap}).\label{fig_fqplot}}
\end{figure}

\begin{figure}
\plotone{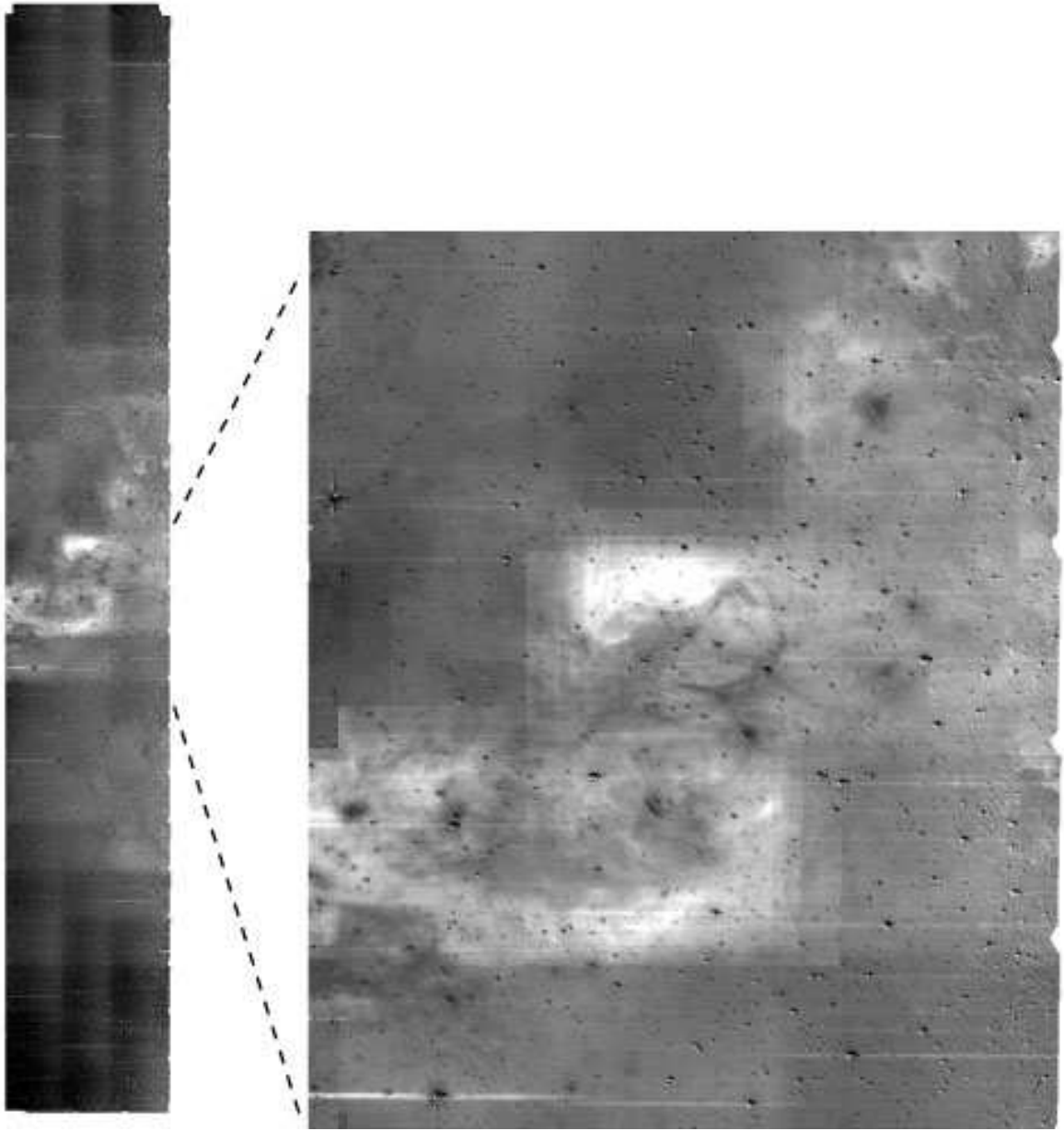}
\caption{Ratio of the standard post-BCD mosaic to our self--calibrated BCD mosaic 
for one AOR strip of the survey (AOR key = 13368320) 
at 8 $\micron$. Display range = [0.8,1.3] MJy sr$^{-1}$.
This ratio highlights the corrections made by the fixed 
and variable offsets derived via self--calibration. Some of 
the sharpest frame boundaries evident here (in the expanded view) are visible in 
the standard post--BCD mosaics, but not in the self--calibrated mosaics. The bright region
around Sgr B1 is at the center of this $0\fdg25 \times 1\fdg75$ strip. 
The lack of banding correction in the post-BCD mosaic causes the 
horizontal streaks at the locations of bright stars.
\label{fig_rat34}}
\end{figure}

The variable detector offset ($F^q$) is illustrated in Figures \ref{fig_fqplot} and \ref{fig_fqmap}.
The time sequence 
of these shows a regular variation each time one of the AOR strips crosses the brightest
part of the Galactic plane. For the InSb detectors used at 3.6 and 4.5 $\micron$,
$|F^q| \lesssim 0.5$ MJy/sr and is usually much smaller than the observed 
sky intensities (which do not include zodiacal light) as shown in Table~1.
The variable offsets can be much larger for the 5.8 and 8 $\micron$ 
detectors, as seen in Fig. \ref{fig_fqplot}. Here the variable 
offsets are correcting for any residual errors in the ``first--frame'' effect (a change in 
detector offset as a function of time since the preceding frame), and more noticeably
for detector droop (where the detector offset changes as a function of the total 
illumination on the detector). 
Figure \ref{fig_rat34} shows the ratio of our mosaic compared to a 
standard post-BCD pipeline mosaic for one strip of our survey (AORID = 13368320) 
at 5.8 and 8 $\micron$. 
This ratio clearly shows artifacts at the frame boundaries, indicating offset errors
in one (or both) of the mosaics. These errors are largest at the bright emission 
through the center of the strip at $b = 0$, especially for the 8 $\micron$ channel.
The facts that (a) the frame seams are faintly visible in the 
post-BCD mosaic and not our self--calibrated versions, and (b) the subsequent science analysis reveals 
no similar discrepancies in the ratios of 5.8 to 8 $\micron$ emission despite independently
derived corrections for each channel, indicate that the variable offsets derived 
from the self--calibration are appropriate. In other words, the self--calibration procedure
does remove artifacts from the images which would otherwise cause systematic errors 
when comparing intensities at different wavelengths.

\subsection{Extended Source Calibration and Zodiacal Light}

The absolute calibration of IRAC is based on the photometry of point sources
in a fixed size aperture. In the present study, we are interested in the surface
brightness of extended emission. Therefore, we have applied extended source corrections
to the data used in this analysis. Specifically, the factors used here are:
[0.944,0.937,0.772,0.737] at 3.6, 4.5, 5.8, and 8 $\micron$ respectively
(Reach et al. 2005; Cohen et al. 2007).

Accurate determination of the colors of diffuse emission also relies on 
having properly subtracted any instrumental offsets. 
The major difficulty in measuring the correct instrumental offset
is that IRAC does not use its shutter to take dark frames. 
Instead, ``skydarks'' (observations at a dark location near 
the north ecliptic pole) are used to in the pipeline 
to remove instrumental offsets. This works well as long as the instrumental offsets 
remain stable between the time that the skydarks are measured and the science data 
are observed. IRAC's 5.8 $\micron$ channel show the greatest instabilities in its
offsets. The lack of constraints that would be provide by true dark frames also
leads to an offset degeneracy in the self--calibration of these data. 
Inspection of Eq. 1 reveals that the procedure is 
insensitive to an arbitrary offset to the derived sky intensity $S^{\alpha}$ as long as
the same offset is subtracted from the derived instrumental offset $F^p$ (or $F^q$).

Finally, for measurement of the intensity and color of diffuse Galactic emission 
the zodiacal light foreground also must be subtracted. The zodiacal light 
must be estimated via a model. The model used for {\it Spitzer} 
is derived from the Kelsall et al. (1998) model of the zodiacal light emission
as observed by the DIRBE instrument on the {\it COBE} spacecraft. The model is 
estimated to be accurate to $\sim 2\%$, however larger unrecognized systematic 
errors may be present. Its use for IRAC data requires interpolation
to IRAC's wavelengths, and, more significantly, extrapolation from the view 
in Earth orbit to the view from {\it Spitzer}'s location several tenths of
an astronomical unit away from Earth. Additionally, because of the skydark subtraction,
the zodiacal light must be known along the line of sight to the north ecliptic
pole as well as toward the Galactic center. Thus the zodiacal light intensity we 
subtract from the mosaic images is the difference between that estimated for 
the Galactic Center observations and that of the skydark observations. 
This is calculated from the mean values of the ZODY$\_$EST and SKYDRKZB keywords 
in the BCD headers, except for the 5.8 $\micron$ channel where these keyword 
values are in error and are inconsistent with the model spectrum; thus the 
estimates from SPOT\footnote{http://ssc.spitzer.caltech.edu/propkit/spot/} are 
used instead for the 5.8 $\micron$ data. 
We do not account for the small gradient in the zodiacal light intensity. 

\begin{figure}
\plotone{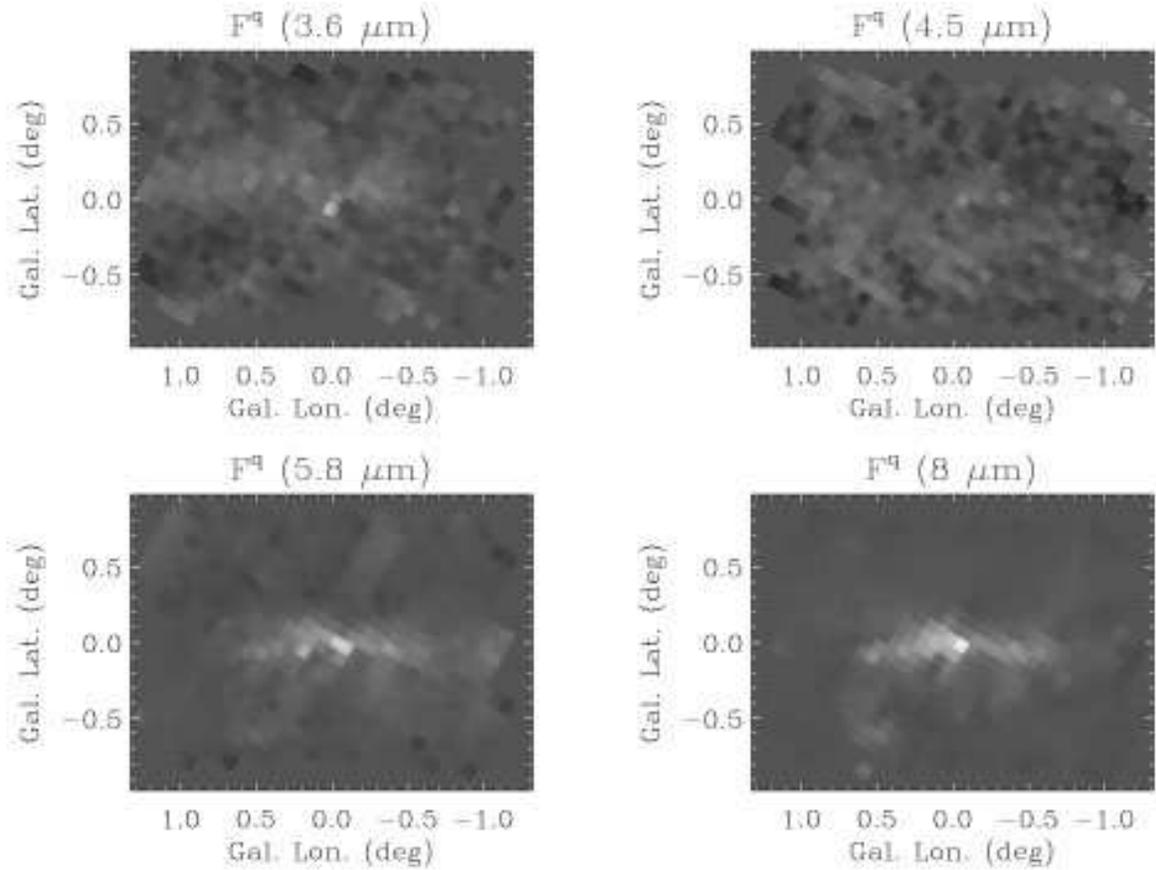}
\caption{Variable offsets per frame ($F^q$) for each channel are mosaicked into images 
in Galactic coordinates. The linear grey scale ranges are the same as shown in 
Figure \ref{fig_fqplot}. 
The 5.8 and 8 $\micron$ channel offsets are relatively large and well--correlated 
with the local sky intensity, and thus these maps resemble very low resolution images
of the Galactic center region.\label{fig_fqmap}}
\end{figure}

Table 1 lists the median and mean intensities of the final GC mosaics, along with the 
self--calibration offsets, and the estimated intensities of the zodiacal light.
The zodiacal light is essentially negligible for the 3.6 and 4.5 $\micron$ data,
but could potentially influence 5.8 and 8 $\micron$ results where intensities are low. 
The variable offset per frame ($F^q$) can reach large values, but 
as revealed in Figure \ref{fig_fqmap},
the largest values are closely confined to Galactic plane where intensities are also 
much higher than the mean or median values.

\clearpage

\begin{deluxetable}{lccccccccccc}
\tabletypesize{\scriptsize}
\tablewidth{0pt}
\tablecaption{Comparison of the Observed Sky Intensities and Offset Terms}
\tablehead{
\colhead{} &
\multicolumn{2}{c}{$I$ (MJy sr$^{-1}$)} & & 
\colhead{$Z$ (MJy sr$^{-1}$)} & & 
\multicolumn{2}{c}{$F^p$ (MJy sr$^{-1}$)} & &
\multicolumn{3}{c}{$F^q$ (MJy sr$^{-1}$)} \\
\cline{2-3}\cline{5-5}\cline{7-8}\cline{10-12}
\colhead{Wavelength ($\micron$)} &
\colhead{median} & 
\colhead{mean} & &
\colhead{mean} & &
\colhead{median} & 
\colhead{mean} & &
\colhead{$\sigma$} & 
\colhead{median} &
\colhead{max}
}
\startdata
3.6 & 14.1 &  23.9 & & 0.097 & & -0.0367 & 0.0173 & & 0.186 & 0.0139  & 1.92  \\
4.5 & 11.0 &  18.5 & & 0.193 & & -0.0613 & 0.0110 & & 0.201 & 0.00906 & 0.874 \\
5.8 & 26.7 &  33.9 & & 1.298 & &  0.0331 & 0.0686 & & 0.930 & 0.0839  & 8.49  \\
8   & 66.1 &  77.1 & & 6.181 & & -0.0663 & -0.315 & & 9.14  & 0.659   & 99.3
\enddata
\tablecomments{$I$ = the observed intensity in the final image. $Z$ = the difference between 
the zodiacal light toward the GC and that toward the skydark calibration field, 
$F^p$ = self--calibration detector offset per detector pixel $p$, $F^q$ = 
self--calibration detector offset per frame $q$.
}
\end{deluxetable}

\clearpage

\end{document}